\documentclass[epj]{svjour}
\usepackage{graphics}
\begin{document}
\title{Nonextensive Nambu Jona-Lasinio Model of QCD matter}
\author{Jacek Ro\.zynek\thanks{\emph{e-mail: jacek.rozynek@ncbj.gov.pl}}
 \and Grzegorz Wilk\thanks{\emph{e-mail: grzegorz.wilk@ncbj.gov.pl}}
}                     
\institute{National Centre for Nuclear Research,
        Department of Fundamental Research, Ho\.za 69, 00-681
        Warsaw, Poland}
\date{Received: date / Revised version: date}

\abstract{We present a thermodynamical analysis of the
nonextensive, QCD-based, Nambu - Jona-Lasinio (NJL) model of
strongly interacting matter in the critical region. It is based on
the nonextensive generalization of the Boltzmann-Gibbs (BG)
statistical mechanics, used in the NJL model, to its nonextensive
version. This can be introduced in different ways, depending on
different possible choices of the form of the corresponding
nonextensive entropies, which are all presented and discussed in
detail. Unlike previous attempts the present approach fulfils the
basic requirements of thermodynamical consistency. The
corresponding results are compared, discussed and confronted with
previous findings.
} 
\maketitle

\section{Introduction}
\label{sec:Introduction}

Some time ago we presented a nonextensive version of the QCD-based
Nambu - Jona-Lasinio (NJL) model of  many-body mean field theory
describing the behavior of strongly interacting matter
\cite{NJL,NJLa,NJL1,NJL2,HK,HK1,HK2,Sousa,Costa}, the $q$-NJL model
\cite{RWJPG,RWConf,RWConf1}. It was based on the nonextensive
generalization of Boltzmann-Gibbs (BG) statistical mechanics used
in the NJL model to its nonextensive version and was characterized
by a dimensionless nonextensivity parameter $q$. At the same time
the other many-body mean field theory of nuclear matter, the
Walecka model \cite{SW,SW1,SW2} formulated on the hadronic level, has also
been generalized to its nonextensive version \cite{Pereira,Pereira1,Pereira2},
similarly as a number of other nonextensive approaches to
different aspects of dense hadronic and quark matter, cf., for
example, \cite{LPQ,LPQ1,LPQ2,LPQ3,LPQ4,GGG,GGG1}.

In short, introduction of nonextensivity to a given model consists
in the replacement of the usual extensive Boltzmann-Gibbs (BG)
distribution by a nonextensive Tsallis distribution \cite{T,T1,T2,T3}
(preserving the original dynamical structure of the
model\footnote{As was done, for example, in the nonextensive
hydrodynamical model discussed in \cite{q-hydro,q-hydro1}.}):
\begin{eqnarray}
\!\!\!\!\!\!\!\!\!\!&& f(X) = C \exp \left( - \frac{X}{T}\right)
\rightarrow f_q(X) =
C_q \exp_q \left(- \frac{X}{T}\right),\label{q-stat_f}\\
 \!\!\!\!\!\!\!\!\!\!&& {\rm where}~\exp_q\left(- \frac{X}{T}\right) = \left[1 -
(1-q)\frac{X}{T}\right]^{\frac{1}{1-q}}, \label{def_expq}
\end{eqnarray}
$X$ denotes the variable of interest, $S$ is the BG-Shannon
entropy and $S_q$ the Tsallis entropy,
\begin{equation}
S = - \sum p_i \ln p_i ~\rightarrow~ S_q = \frac{\sum \left( 1 -
p_i^q\right)}{q -1} \label{q-stat_S}
\end{equation}
(for $q \rightarrow 1 $ both entropies coincide).

However, one should be aware of the fact that the nonextensive
distribution (\ref{q-stat_f}) can also be regarded as an example
of a quasi-power law extrapolating between a pure, scale
invariant, power-like form for large $X$ and an exponential one
for small $X$. In fact, it is observed in all branches of physics
\cite{T}, including multiparticle production experiments
\cite{multipart,multiparta,qMB,multipart1,multipart1a,multipartB,mpo,mpo1,mpo2,mpo3,mpo4,mpo5,CW,CWa},
studies of complex systems \cite{qBiro,qBiro1,qBiro2,qBiro3,qBiro4} or systems on lattices
\cite{Birolat,Birolat1,Birolat2,Birolat3}. As such, it can be derived in a variety of ways,
not necessarily connected with statistical mechanics and not based
on any entropy \cite{WqW1,WqW1a,WqW1b,WqW2,qQCD,qQCD1,qBiro,qBiro1,qBiro2,qBiro3,qBiro4}. Because all our
further considerations are based on  nonextensive thermodynamics
with some nonextensive form of entropy, it should be mentioned
that such an approach is fully compatible with the usual
traditional extensive thermodynamics \cite{M,M1,M2,M3,M4,M5}.

Our motivation behind investigating the possible effects of
introducing a nonextensive environment (with $q$ not equal to
unity) into the original extensive mean field approach of the NJL
model remains the same as before \cite{RWJPG,RWConf,RWConf1}. Referring
there for details, basically it is a desire to account, in a
phenomenological way, for the numerous factors which make the
assumptions of the mean field approach questionable. For example,
it does not account for possible intrinsic (nonstatistical)
fluctuations and correlations (especially short range ones), which
are common when quark matter is produced in high energy heavy ion
collisions. In such collisions it emerges in small and rapidly
evolving samples, the spatial configuration of which is far from
being uniform, and no global equilibrium is established
\cite{departure,departure1,departure2,departure3}. The NJL model is therefore usually supplemented
by some additional dynamical ingredients (like, for example, the
Polyakov loops (cf., \cite{PLoop,PLoop1,PLoop2}) or by some retardation effects
\cite{HK,HK1,HK2}). Instead of these, we propose to change summarily the
type of environment by allowing it to be nonextensive, with the
degree of nonextensivity provided by $q - 1$ and without
specifying its dynamical origins. On the other hand, it is
expected that addition to such an approach any dynamical
ingredient of the type mentioned above should result in a
diminishing of the value of $|q - 1|$ used. In fact, a fitted
value of $q = 1$ would signal that there is no need for any
further nonextensivity based on nonextensive statistical
mechanics\footnote{Such a situation was encountered some time ago
in multiparticle production processes \cite{qMB}. The gradual
accounting for the intrinsic dynamical fluctuations in the
hadronizing system by switching from a simple nonextensive
distribution of the type of Eq. (\ref{q-stat_f}) to some dynamical
formula with temperature fluctuations, substantially lowered the
value of $|q - 1|$.}. Its description in terms of nonextensive
statistical mechanics was replaced by some direct, dynamical one.

Unlike the simple situations leading to Eq. (\ref{q-stat_f}), in
the $q$-NJL model one deals with fermions and has to account for
both particles and antiparticles with nonzero chemical potentials.
In such cases the formulation of the nonextensive formalism is not
unique, especially when one wishes to fulfil the basic
requirements of thermodynamical consistency \cite{JR}. First of
all, the proper number of constituents in this case is not given
by the usual nonextensive particle occupation numbers, $n_q$, but
rather by their $q$ powers, $n_q^q$
\cite{PPMU,TPM,TPM1,consistency,CW1,CW1a}. Further, the nonextensivity
parameter $q$ is not necessarily constant (actually this was the
point overlooked in our previous work \cite{RWJPG,RWConf,RWConf1}). As
shown in \cite{JR} the parameter $q$ depends on whether one
describes particles or antiparticles, (changing, for example, from
$q$ for particles to $2 - q$ for antiparticles), or, in other
cases, it can depend on the value of the density $n_q$ (by
changing, again, from $q$ to $2 - q$, but this time at the Fermi
level at which $n_q = 1/2$).

The effective number of constituents $n_q^q$ was already used in
\cite{LPQ,LPQ1,LPQ2,LPQ3,LPQ4,LP,LPSTAR} (with different choices of the form of
nonextensive densities) and also in a recent new version of the
$q$-Walecka model proposed in \cite{Santos}. However, the
necessity of using in some cases for the exponent both $q$ and $2
- q$ was fully explored in the analysis of dense hadronic matter
only recently \cite{DeppmanJ,AD,AD1,AD2,AD3,AD4,AD5,AD6}. In this respect our previous
calculations \cite{RWJPG,RWConf,RWConf1}, performed assuming that the
exponent $q$ is always constant, were not correct. Therefore, we
shall repeat them here, this time for all plausible choices of
densities $n_q$ and for the correct choices of the corresponding
$n_q^q$ as proposed recently in \cite{JR}. At the same time the
physical meaning of using the $q > 1$ or $q < 1$ type of
nonextensive environment (usually connected with, respectively,
intrinsic dynamical fluctuations \cite{qfluct,qfluct1,qfluct2} or correlations
\cite{qcorrel,qcorrel1,qcorrel2}) will also be discussed.

The outline of our work is as follows. After providing in Section
\ref{sec:RNJL} a short reminder of the NJL model used, we present
in Section \ref{sec:q-NJL} the revisited version of our $q$-NJL
model. Different ways of implementing  it will be presented here
and discussed in detail. Section \ref{sec:Results} contains our
results with special emphasis on the presentation and detailed
discussion of the different forms of nonextensive entropy used and
on their interrelations. In particular, we show that the usual
ordering of entropies  with respect to the value of the
nonextensivity parameter $q$, $S(q<1) > S(q=1) > S(q>1)$ \cite{T,T1,T2,T3},
in the case of $q$-NJL, with $q$-dependent occupation numbers, can
be reversed. We shall also present here results  on the
susceptibilities and the heat capacity of nonextensive nuclear
matter. Section \ref{sec:Summary} summarizes our presentation.

\section{Reminder of the NJL model used}
\label{sec:RNJL}

With regard to the NJL model used by us, we follow the version
presented in \cite{Sousa}, with an effective lagrangian suitable
for the bosonization procedure and with a four quark interaction
only given by (for more details see \cite{RWJPG}):
\begin{eqnarray}
{\cal L}_{eff} &=& \bar{q}\left(i {\gamma}^{\mu}\partial_\mu
-\hat{m}\right)q\! +\! S_{ab}\left[\left(\bar{q}\lambda^a q
\right)\left( \bar{q}\lambda^b q \right)\right] + \nonumber\\
&& +\, P_{ab}\left[\left(\bar{q} i\gamma_5 \lambda^a q
\right)\left( \bar{q} i\gamma_5\lambda^b q \right)\right],
\label{lagr_eff}
\end{eqnarray}
where $\hat{m} = {\rm diag}\left(m_u, m_d, m_s\right)$ and
\begin{eqnarray}
S_{ab} &=& g_S \delta_{ab} + g_D D_{abc}\left\langle \bar{q}
\lambda^c
q\right\rangle, \label{sab}\\
P_{ab} &=& g_S \delta_{ab} - g_D D_{abc}\left\langle \bar{q}
\lambda^c q\right\rangle. \label{pab}
\end{eqnarray}
It is invariant under the chiral SU$_L(3)\otimes$SU$_R(3)$
transformations described by coupling constant $g_S$ (except for
the current quarks mass term) and contains a term breaking the
U$_A(1)$ symmetry (described by thecoupling constant $g_D$), which
reflects the axial anomaly in QCD. $D_{abc}$ are the SU(3)
structure constants $d_{abc}$ for $a,b,c=(1,2,\dots,8)$ whereas
$D_{0ab} = -\delta_{ab}/\sqrt{6}$ and $D_{000} = \sqrt{2/3}$. We
work with $q = (u,d,s)$ quark fields with three flavors, $N_f =
3$, and three colors, $N_c = 3$, $\lambda^a$ are the Gell-Mann
matrices, $a = 0,1,\ldots , 8$ and ${\lambda^0=\sqrt{2/3} \, {\bf
I}}$.

Integrating over the momenta of quark fields in the functional
integral with ${\cal L}_{eff}$ one obtains an effective action
expressed in terms of $\sigma$ and $\varphi $, the natural degrees
of freedom of low energy QCD in the mesonic sector ($\mbox{Tr}$
stands for taking the trace over indices $N_f$ and $N_c$):
\begin{eqnarray}
W_{eff}[\varphi,\sigma] &=& -\frac{1}{2}\left(
\sigma^{a}S_{ab}^{-1} \sigma^{b}\right)  - \frac{1}{2}\left(
\varphi^{a}P_{ab}^{-1}\varphi
^{b}\right) - \label{action}\\
&-& i\mbox{Tr}\,\mbox{ln}\Bigl[i\gamma^{\mu}\partial_{\mu}-\hat{m}%
+\sigma_{a}\lambda^{a}+(i\gamma_{5})(\varphi_{a}\lambda^{a})\Bigr]\,.
\nonumber
\end{eqnarray}
The first variation of $W_{eff}$ results in the gap equations for
the constituent quark masses $M_i$:
\begin{eqnarray}
 M_i = m_i - 2g_{_S} \big <\bar{q_i}q_i \big > -2g_{_D}\big
 <\bar{q_j}q_j\big > \big <\bar{q_k}q_k \big >, \label{gap}
 \end{eqnarray}
with cyclic permutation of $i,j,k =u,d,s$ and with the quark
condensates
\begin{equation}
\big <\bar{q}_i q_i \big > = -i \mbox{Tr}[ S_i(p)] \label{qcond}
\end{equation}
($S_i(p)$ is the quark Green function, $m_i$ is the current mass
of quark of flavor $i$; notice that nonzero $g_D$ introduces
mixing between different flavors).

We consider a system of volume $V$, temperature $T$ and $i^{th}$
quark chemical potential $\mu_i$ characterized by the baryonic
thermodynamic potential of the grand canonical ensemble,
\begin{equation}
\Omega (T, V, \mu_i )= E- TS - \sum_{i=u,d,s} \mu _{i} N_{i},
\label{tpot}
\end{equation}
with quark density equal to $\rho_i = N_i/V$, baryonic chemical
potential $\mu_B= \frac{1}{3} (\mu_u+\mu_d+\mu_s)$ and  baryonic
matter density $\rho_B = \frac{1}{3}(\rho_u+\rho_d+\rho_s)$. The
internal energy, $E$, the entropy, $S$, and the particle number,
$N_i$, are given by \cite{Sousa,Costa} (here $E_i = \sqrt{M_i^2 +
p^2}$):
\begin{eqnarray}
\!\!\!\!\!\!\!E &=&- \frac{ N_c}{\pi^2} V\sum_{i=u,d,s}\left[
   \int p^2 dp  \frac{p^2 + m_{i} M_{i}}{E_{i}}
   \left( 1 - n_{i}- \bar{n}_{i} \right) \right] - \nonumber\\
   && - g_{S} V \sum_{i=u,d,s}\, \left( \langle \bar{q}_{i}q_{i}\rangle \right)^{2}
   - 2 g_{D}V \langle \bar{u}u\rangle \langle \bar{d}d\rangle \langle \bar{s}s\rangle , \label{energy} \\
 S &=& \! -\frac{ N_c}{\pi^2} V \sum_{i=u,d,s}\int p^2 dp \cdot
 \tilde{S}, \label{entropy}\\
 \tilde{S} &=&  \bigl[ n_{i} \ln n_{i}+(1-n_{i})\ln (1-n_{i})
   \bigr]\!\! +\!\! \bigl[ n_{i}\rightarrow 1 - \bar n_{i}
   \bigr],\label{Barentropy}\\
N_i &=& \frac{ N_c}{\pi^2} V \int p^2 dp
  \left( n_{i}-\bar n_{i} \right) \label{number}.
\end{eqnarray}

All these quantities depend on the quark and antiquark occupation
numbers, $n_i$ and $\bar n_i$. They can be obtained in different
ways; we shall use the standard Jayne's information-theoretic
approach and derive them from a given measure of information
represented by the entropy functional (\ref{Barentropy}).
Extremalizing it under constraints imposed by the total number of
particles, $\hat{N}$, and total energy of the system, $\hat{E}$
($E_i$ is the energy of the $i$-th energy level) \cite{EM,EM1,PPMU},
\begin{equation}
\sum_{i} \left(n_i - \bar{n}_i\right) = \hat{N}\quad{\rm and}\quad
\sum_i \left(n_i + \bar{n}_i \right) E_i = \hat{E},
\label{eq:constraints}
\end{equation}
one finds that ($e(x) = \exp (x)$):
\begin{eqnarray}
&&n_{i} = \frac{1}{ e \left( x_i\right) + 1},\qquad \quad \bar
n_{i} = \frac{1}{e \left( \bar{x}_i \right)
+ 1}, \label{FD}\\
&& x_i = \beta ( E_i - \mu),\qquad \quad\bar{x}_i = \beta ( E_i +
\mu). \label{xbarx}
\end{eqnarray}

The values of the quark condensates in Eq. (\ref{gap}) are now
given by
\begin{eqnarray}
\big <\bar{q}_i  q_i \big> \!= \!\ - \frac{ N_c}{\pi^2} \!\!\!
\sum_{i=u,d,s}\left[ \int \frac{p^2M_i}{E_i} (1\,-\,n_{i}-\bar
n_{i})\right]dp .\label{gap1}
\end{eqnarray}
Eqs. (\ref{gap}) and (\ref{gap1}) form a self consistent set of
equations to calculate, for a given $T$ and $\mu$, the effective
quark masses $M_i$ and values of the corresponding quark
condensates, $\langle \bar{q}_iq_i\rangle$, and, further, all
quantities of interest. With these effective masses $M_i$ the
occupation numbers (\ref{FD}), the form of which was obtained
assuming a noninteracting Fermi gas, will get their dynamical
input defined by the form of lagrangian used, Eqs.
(\ref{lagr_eff}) and (\ref{action}). The model is fixed by the
coupling constants $g_S$ and $g_D$ and by the cutoff in
three-momentum space $\Lambda$, which is used to regularize the
momentum space integrals and the current quark masses $m_i$. The
values of all parameters are the same as in \cite{RWJPG}.

\section{The $q$-NJL model revisited}
\label{sec:q-NJL}

\subsection{The problem of thermodynamic consistency}
\label{sec:thC}

Thermodynamical consistency means that one has to preserve the
standard thermodynamical relationships among thermodynamical
variables such as, for example, entropy, energy and temperature,
\begin{equation}
\left(\frac{\partial S}{\partial E}\right)_{\rho=N/V} =
\frac{1}{T}.\label{thercons}
\end{equation}
As shown in \cite{MEnt_cons,MEnt_cons1}, any thermostatistical formalism
constructed by following Jayne's maximum entropy prescription
complies with the thermodynamical relationships. A
thermodynamically consistent formulation is then obtained by the
appropriate identification of the relevant constraints with the
extensive thermodynamical quantities (like number of particles $N$
or energy $E$) and by identification of the corresponding Lagrange
multipliers with the appropriate intensive thermodynamical
quantities (like temperature $T$ and chemical potential $\mu$). In
our case we shall extremalize some appropriately chosen entropic
measures using some specifically chosen constraints. Because this
procedure is not unique, we shall discuss three different versions
of the $q$-NJL model, based on different choices of entropic
measure $\tilde{S}_q$ proposed in \cite{JR}.

\subsection{Approach with Tsallis' cut-off prescription for $q > 1$}
\label{sec:a}

For $ q > 1$ the straightforward approach uses a
$q$-generalization of the entropic measure $\tilde{S}$ in  Eq.
(\ref{Barentropy}), which reduces to it for $q \rightarrow 1$,
\begin{eqnarray}
\tilde{S}^{(a)}_q &=& \sum_i \left[ n^q_{qi} \ln_q n_{qi} +
(1-n_{qi})^q\ln_q
(1-n_{qi}) \right] + \nonumber\\
&& + \left[ n_{qi} \rightarrow 1 - \bar{n}_{qi}\right], \label{qBarent}\\
&& {\rm where}\quad \ln_q x = \frac{x^{1-q} - 1}{1 - q}.
\label{qlog}
\end{eqnarray}
To fulfil the basic requirements of thermodynamical consistency
with such a form of entropy functional, the constraints must have
the following form \cite{JR}:
\begin{eqnarray}
\sum_i \left( n_{qi}^q - \bar{n}_{qi}^q\right) &=& \hat{N},\quad
\sum_i \left( n_{qi}^q + \bar{n}_{qi}^q\right)E_{qi} = \hat{E}.
\label{qconstraints}
\end{eqnarray}
Such a form reflects the fact that in the nonextensive environment
parameterized by $q$ the relevant number of constituents is given
not by $n_q$, but by $n_q^q$. This is because, as shown in
\cite{qBiro,qBiro1,qBiro2,qBiro3,qBiro4}, the occurrence of the nonextensivity can be formally
understood as a result of the action of some effective
interactions between constituents of the system, which therefore
lose their independence. The strength of these interactions is
proportional to $ \zeta = q-1$: they are attractive for $\zeta <
0$ and repulsive for $\zeta > 0$, and they are not accounted for
by the usual description of the system considered; for example, in
our case of the $q$-NJL model they act on top of the interaction
described by the lagrangian defined by Eq. (\ref{lagr_eff}).

Extremalization of $\tilde{S}_q^{(a)}$ under constraints
(\ref{qconstraints}) results
 in the following quark and antiquark occupation numbers\footnote{In what
follows we shall use the notation $n_{\hat{q}}$ or $n_{\tilde{q}}$
(and $e_{\hat{q}}$ or $e_{\tilde{q}}$) in situations where $q$ in
the above definitions will be replaced by a suitably defined
$\hat{q}$ or $\tilde{q}$. In other cases the $q$ in $X_q$ are just
subscripts indicating the nonextensive origin of variable $X$.}:
\begin{eqnarray}
n_{qi}\! &=&\! \frac{1}{e_q\left( x_{qi} \right) + 1},\qquad
\bar{n}_{qi}\! =\! \frac{1}{e_q\left( \bar{x}_{qi}\right)
 + 1}, \label{qFD}\\
{\rm where}&& x_{qi} = \beta\left( E_{qi} - \mu\right),\quad
\bar{x}_{qi} = \beta\left( E_{qi} + \mu\right),\label{qxbarx}\\
{\rm and}~~&& e_q (x)  = [ 1 + (q-1) x ]^{\frac{1}{q-1}}.
\label{expq}
\end{eqnarray}

As in the extensive case, Eq. (\ref{FD}), the above effective
occupation numbers are derived for a Fermi gas, i.e., for a
situation with no interactions. However, this time it is immersed
in a nonextensive environment, which, as mentioned above,
introduces some effective interaction \cite{qBiro,qBiro1,qBiro2,qBiro3,qBiro4}. Therefore, it
is in reality an assumption that we shall use them in further
calculations where $E_{qi} = \sqrt{M_{qi} + p^2}$ and $M_{qi}$
denotes the quark mass obtained from the $q$-version of the gap
equation (\ref{gap}), cf. Eq. (\ref{qgap}) below. In addition to
the dynamical input mentioned before when discussing Eq.
(\ref{FD}), this is additional source of nonextensivity in a
$q$-NJL model. For $q\rightarrow 1$ we recover Eqs. (\ref{FD}),
(\ref{gap}) and (\ref{gap1}) of the usual NJL model. The same
remark applies to the other two approaches presented
below\footnote{Note that for $T\rightarrow0$ one always gets
$n_q(\mu,T)\rightarrow n(\mu,T)$, irrespective of the value of
$q$. This means that we can expect a nonextensive signature only
for high enough temperatures. A similar situation is encountered
in models using a Polyakov-loop potential (for example, in a
Polyakov-loop extended quark meson model \cite{PolLoop}).}.

Definition (\ref{expq}) must be supplemented by a condition
ensuring that thefunction $e_q(x)$ is always non-negative real
valued (known as Tsallis' cut-off prescription),
\begin{eqnarray}
e_q(x) &=& 0\quad {\rm for}\quad [ 1 + (q-1)x ] \leq 0.
\label{expqg1}\\
{\rm or}\quad n_{qi} &=& 1 \quad {\rm for}\quad E_{qi} < \mu -
\frac{1}{\beta(q-1)}. \label{limqg1p}
\end{eqnarray}
There are no limitations on the distribution of antiparticles,
$\bar{n}_{qi}$. We observe always that $n_{q>1} (x)
> n_{q=1}(x)$ (cf. also \cite{TPM,TPM1}).

\subsection{Approach with Tsallis' cut-off prescription for $q < 1$}
\label{sec:b}

The case where $q > 1$ is the usual range explored in all previous
investigations of this type, cf. \cite{Pereira,Pereira1,Pereira2,LPQ,LPQ1,LPQ2,LPQ3,LPQ4,LP,LPSTAR}).
However, because in the $q$-NJL model we must consider on the same
footing both particles and antiparticles, we have to address how
to formulate our approach for the $ q < 1$ case. It is easy to
check that using the previous description we would have to replace
in this case the condition (\ref{expqg1}) by
\begin{eqnarray}
e_q(x) &=& +\infty \quad {\rm for}\quad [ 1 + (q-1)x
] \leq 0.\label{expql1}\\
{\rm or~}\quad n_{qi} &=& 0 \quad {\rm for}\quad E_{qi} >
\mu + \frac{1}{\beta(1-q)},\label{limql1p}\\
{\rm and}\quad \bar{n}_{qi} &=& 0 \quad {\rm for}\quad E_{qi} > -
\mu + \frac{1}{\beta(1-q)}.\label{limql1a}
\end{eqnarray}
However, this means that for reasonable temperatures $T$ the phase
space for antiparticles in such an approach would be practically
completely cut-off.

Therefore, if we intend to extend this approach to the $q$-NJL
model and consider the case $q<1$, we have to modify the form of
the entropy functional and replace $\tilde{S}^{(a)}$ by \cite{JR}
\begin{eqnarray}
\tilde{S}^{(b)}_q &=& \sum_i \left[ n^q_{qi} \ln_q n_{qi} +
(1-n_{qi})^q\ln_q
(1-n_{qi}) \right] + \nonumber\\
&& + \left[ n_{qi} \rightarrow 1 - \bar{n}_{\hat{q}i};~q
\rightarrow \hat{q} = 2 - q \right] \label{qBarenta},
\end{eqnarray}
with accordingly modified constraints,
\begin{eqnarray}
\sum_i \left( n_{qi}^q - \bar{n}_{\hat{q}i}^{\hat{q}}\right) &=&
\hat{N},\quad \sum_i \left( n_{qi}^q +
\bar{n}_{\hat{q}i}^{\hat{q}}\right)E_{qi} = \hat{E}.
\label{mqconstraints}
\end{eqnarray}
The resulting occupation numbers are still given by Eq.
(\ref{qFD}) with nonextensive parameter $q$ for particles and with
$q$ replaced by its dual, $\hat{q} = 2 -q$, for antiparticles.
This means that in $\bar{n}_{\hat{q}i}$ one now has
\begin{eqnarray}
e_q\left(\bar{x}_{qi}\right) \rightarrow  e_{\hat{q}}\left(
\bar{x}_{qi}\right) &=& \left[ 1 + \left( \hat{q} -
1\right)\bar{x}_{qi}\right]^{\frac{1}{\hat{q} - 1}}.\label{eqa}
\end{eqnarray}
In this case  we always observe that $n_{q<1}(x) < n_{q=1}(x)$.

With such a choice of nonextensive parameter there are no longer
any problems for antiparticles with regard to phase space
limitations and their physical meaning. However, the price one
pays is that now particles and antiparticles are described by
different, albeit connected, nonextensive parameters: $q$ for
particles and its dual, $\hat{q} = 2 - q$, for antiparticles.
Therefore the choice $q < 1$ for particles means $\hat{q}
> 1$ for antiparticles (i.e., it remains the same as in Section \ref{sec:a} above).
The possible justification of such a choice could be that
antiparticles, in contrast to particles, do not have a Fermi level
(in the sense that their states are not fully occupied).
Therefore, they are not so strongly correlated as particles, but
rather experience dynamical fluctuations (believed to be described
by $q > 1$ \cite{qfluct,qfluct1,qfluct2}).

\subsection{Approach without Tsallis' cut-off prescription}
\label{sec:c}

The Tsallis cut-off prescriptions needed in the above two choices
of the entropy functional, $\tilde{S}_q^{(a)}$ in Eqs.
(\ref{qBarent}) and $\tilde{S}_q^{(b)}$ in Eq. (\ref{qBarenta}),
limit considerably the allowed phase space and, concerning the
$q$-NJL model, they apparently do not have any physical
justification. Therefore, in \cite{TPM,TPM1,consistency} an alternative
approach was proposed eliminating the cut-off prescription by a
suitable choice of the entropy functional. Thus far it has been
used in \cite{CW1,CW1a,Santos,DeppmanJ} for particles only. In the case
of the $q$-NJL model where both particles and antiparticles have
to be considered together this approach has to be modified. The
proposed choice can be summarized as follows \cite{JR}:
\begin{itemize}
\item for fermions above the Fermi sea level, i.e. for $x_{qi}
> 0$ (or $n_{qi} \in \left[ 0, 1/2\right]$), we always use $q > 1$;

\item for fermions below the Fermi sea level, i.e. for $x_{qi} <
0$ (or $n_{qi} \in \left( \frac{1}{2}, 1\right]$), we always use
$\hat{q} = 2 - q < 1$;

\item for antifermions we assume $q > 1$ over the whole region of
$x$ (or for $n_{qi} \in [0,1]$).
\end{itemize}
These requirements correspond to the following, third, choice of
the corresponding $q$-entropy functional:
\begin{eqnarray}
\tilde{S}_q^{(c)} &=& \sum_i  \left[ {n}^{\tilde{q}}_{\tilde{q}i}
\ln_{\tilde{q}} {n}_{\tilde{q}i} +
(1 - {n}_{\tilde{q}i})^{\tilde{q}}\ln_{\tilde{q}} (1 - {n}_{\tilde{q}i}) \right] + \nonumber\\
 &&+ \left[ {n}_{\tilde{q}i} \rightarrow \bar{n}_{qi},~ \tilde{q} \rightarrow q \right] \label{SSq}
\end{eqnarray}
and constraints in the following form:
\begin{eqnarray}
\!\!\!\! \sum_i \left( {n}_{\tilde{q}i}^{\tilde{q}} -
\bar{n}_{qi}^q\right) &=& \hat{N},\quad \sum_i \left(
{n}_{\tilde{q}i}^{\tilde{q}} + \bar{n}_{qi}^q\right) E_{qi} =
\hat{E}. \label{qconstraints1}
\end{eqnarray}
where $\tilde{q}$ now changes at the Fermi surface in the
following way:
\begin{eqnarray}
\tilde{q} &=& \left\{
\begin{array}{ll}
q\quad & \quad {\rm for}\quad x_{qi} \ge 0,\\
& \\
\hat{q} = 2 - q & \quad {\rm for}\quad x_{qi}< 0.\\
\end{array} \right.\label{hattilde}
\end{eqnarray}
With such choices one has the following occupation numbers:
\begin{eqnarray}
\hspace{-2cm} {\rm for~particles:}\qquad\quad n_{\tilde{q}i} &=&
\frac{1}{e_{\tilde{q}}\left(x_{qi}\right) + 1}, \label{nq3}\\
\hspace{-2cm}{\rm for~antiparticles:}\quad\quad \bar{n}_{qi} &=&
\frac{1}{e_q\left( \bar{x}_{qi}\right) + 1} \label{nq3a}
\end{eqnarray}
where $e_{\tilde{q}}(x) = \left[1 + \left(\tilde{q} -
1\right)x\right]^{\frac{1}{\tilde{q} - 1}}$ with $\tilde{q}$ given
by Eq. (\ref{hattilde}), $e_q(x)$ is defined in Eq. (\ref{expq})
whereas $x = x_{qi}$ and $\bar{x}_{qi}$ are defined in Eq.
(\ref{qxbarx}).

In this case, instead of being either always greater (for $q
>1$, as in case $(a)$) or smaller (for $q < 1$, as in case $(b)$)
than the respective values for the extensive case, we observe that
$n_{\tilde{q}} < n_{q=1}$ below the Fermi surface (i.e., for $x <
0$) and $n_{\tilde{q}} > n_{q=1}$ above the Fermi surface (cf.
also \cite{TPM,TPM1}.

The most important observation which must be kept in mind is that
whereas in this prescription $n_{\tilde{q}i}$ has smooth behavior
for all values of $x_{qi}$, the $\tilde{q}$ (\ref{hattilde}) and
therefore also $n_{\tilde{q}i}^{\tilde{q}}$ have a jump at
$x_{qi}=0$, i.e., for $n_{qi} = 1/2$ (at Fermi surface), the
physical meaning of which is so far unclear.

\subsection{Problem of particle-hole symmetry in a nonextensive
environment} \label{sec:particle-hole}

In an extensive situation (at least in the mean field
approximation), with occupation numbers given by Eq. (\ref{FD}),
the particle-hole symmetry is preserved,
\begin{equation}
n(x) + n(-x) = 1, \label{phsymm}
\end{equation}
and one can interpret holes among the negative energy states as
anti-particles with the corresponding positive energy.

However, any short range interaction violates relation
(\ref{phsymm}) by enhancing the tail of the distribution above the
Fermi level \cite{AHRS}. In a nonextensive environment this
symmetry (\ref{phsymm}) is violated and holds only in a dual
fashion \cite{BSZ}. For example, when implementing nonextensivity
as in Sections \ref{sec:a} and \ref{sec:b}, we observe that
\begin{equation}
n_q(x) + n_{\hat{q}}(-x) = 1. \label{no-phsymm}
\end{equation}
On the other hand, the prescription described in Section
\ref{sec:c} apparently restores the particle-hole symmetry,
because now
\begin{equation}
n_{\tilde{q}}(x) + n_{\tilde{q}}(-x) = 1. \label{phsat}
\end{equation}
However, one should keep in mind that, as was stressed before, the
relevant momentum distributions of particles are now given by
$n_q^q$, not by $n_q$, because one deals not with free particles
but with particles embedded in nonextensive environment
\cite{qBiro,qBiro1,qBiro2,qBiro3,qBiro4}. It means therefore that in our case this symmetry
does not work.

\subsection{Basic formulas of the $q$-NJL model}
\label{sec:basic}

The basic quantity is now the $q$-version of the grand canonical
potential (\ref{tpot}),
\begin{equation}
\Omega_q (T, V, \mu_i )= E_q - TS_q - \sum_{i=u,d,s} \mu _{i}
N_{qi}.  \label{qtpot}
\end{equation}
The pressure and the energy density are defined as, respectively,
\begin{eqnarray}
P_q(\mu,T) &=& - \frac{1}{V}[ \Omega_q(\mu,T) - \Omega_q(0,0)],\label{pressure}\\
\varepsilon_q(\mu,T) &=& \frac{1}{V} [E_q(\mu,T) - E_q(0,0)],
\label{edensity}
\end{eqnarray}
where $\Omega_q(0,0) = E_q(0,0)$ denotes the vacuum energy.

The first derivatives give, respectively, $q$-entropy and
$q$-density,
\begin{equation} S_q = \sum_{i=u,d,s} \frac{\partial
\Omega_q}{\partial T}\Big|_{\mu}\quad {\rm and}\quad \varrho_q =
\sum_{i=u,d,s} \frac{\partial \Omega_q}{\partial \mu}\Big|_T .
\label{SRho}
\end{equation}
The second derivatives result in nonextensive versions of the heat
capacity, $C_{\mu}$, and the barionic susceptibility $\chi_B$,
\begin{equation}
\quad C_{\mu} = \frac{\partial S_q}{\partial
T}\Big|_{\mu}\quad{\rm and}\quad \chi_B = \frac{\partial
\varrho_q}{\partial \mu}\Big|_T.\label{ChiC}
\end{equation}
Because
\begin{eqnarray}
\!\!\!\!\!\!\!dE_q =  \frac{\partial E_q}{\partial
S_q}dS_q+\frac{\partial E_q}{\partial N_q}dN_q = \frac{\partial
E_q}{\partial T}dT + \frac{\partial E_q}{\partial \mu}d\mu,
\label{dEq}
\end{eqnarray}
one can define temperature the $T$ and the heat capacity $C_{\mu}$
as, respectively,
\begin{equation}
\frac{1}{T} = \frac{\partial S_q}{\partial E_q}\quad {\rm
and}\quad \frac{1}{T^2 C_{\mu}} = - \frac{\partial^2 S_q}{\partial
E_q^2}. \label{T_and_C}
\end{equation}

The corresponding nonextensive energy, $E_q$, entropy, $S_q$, and
number density, $N_q$, are now given by
\begin{eqnarray}
\!\!\!\!\!\!\!\!\! E_q &=& - \frac{ N_c}{\pi^2}
V\!\!\!\sum_{i=u,d,s}\left[
   \int p^2 dp  \frac{p^2 + m_{i} M_{qi}}{E_{qi}}
   (1 - n^q_{qi}- \bar{n}^q_{qi}) \right] - \nonumber\\
   &-& g_{S} V\!\! \sum_{i=u,d,s}\!\! \left(\big <
\bar{q}_{i}q_{i}\big >_q \right)^{2}
   - 2 g_{D}V \big < \bar{u}u\big >_q \big < \bar{d}d\big >_q \big <
\bar{s}s\big >_q ; \label{q_energy}\\
 S_q &=& -\frac{ N_c}{\pi^2} V \sum_{i=u,d,s}\int p^2 dp \cdot
 \tilde{S}^{(R)}_{qi} \label{qEnt}\\
N_{qi} &=& \frac{ N_c}{\pi^2} V \int p^2 dp
  \left( n_{qi}^q-\bar n_{qi}^q \right). \label{qnumber}
\end{eqnarray}
The $q$-version of the gap equation (\ref{gap}) for the
constituent quark mass in the nonextensive environment is now
\begin{eqnarray}
 M_{qi} = m_i - 2g_{_S} \big <\bar{q_i}q_i \big >_q -2g_{_D}\big
 <\bar{q_j}q_j\big > \big <\bar{q_k}q_k \big >_q, \label{qgap}
 \end{eqnarray}
with a $q$-dependent quark condensate, which, following
\cite{LPQ,LPQ1,LPQ2,LPQ3,LPQ4,LP,LPSTAR}, is adopted in the form:
\begin{equation}
\langle \bar{q}_iq_i \rangle_q  =  - \frac{ N_c}{\pi^2}\!
\sum_{i=u,d,s}\!\! \left[ \int \frac{p^2M_{qi}}{E_{qi}}
(1\,-\,n^q_{qi}\!-\! \bar{n}^q_{qi})\right]dp. \label{q_gap}
\end{equation}
Depending now on the form of  the entropic functional
$\tilde{S}^{(R)}_q$ used, there are the following three
possibilities for the nonextensivity parameters.

The first choice is that for $q > 1$
\begin{equation}
\tilde{S}_q^{(R)} = \tilde{S}_q^{(a)}, \label{R=a}
\end{equation}
as given by Eq. (\ref{qBarent}) supplemented by constraints
(\ref{qconstraints}) in which the parameter $q$ is constant and
remains the same for particles and antiparticles. As a result, the
$n_{qi}$ and $\bar{n}_{qi}$ are given by Eq. (\ref{qFD}), and the
effective distributions, $n_{qi}^q$ and $\bar{n}_{qi}^q$, have the
same value of the power index $q$ for all values of $n_{qi}$ and
$\bar{n}_{qi}$.

The second choice, to be used if one intends to investigate the
case of $q < 1$, is
\begin{equation}
\tilde{S}_q^{(R)}=\tilde{S}_q^{(b)}, \label{R=b}
\end{equation}
as given by Eq.(\ref{qBarenta}) with constrains
(\ref{mqconstraints}). In this case the parameter $q$ for
particles is replaced for antiparticles by its dual, $\hat{q} = 2
- q$. As a result, whereas formally $n_{qi}$ and $\bar{n}_{qi}$
are again given by Eq. (\ref{qFD}), the parameter $q$ for
particles is replaced by $\hat{q}$ for antiparticles. The same
situation is encountered for the effective distributions which are
now equal to, respectively, $n^q_{qi}$ and
$\bar{n}^{\hat{q}}_{\hat{q}i}$.

The third choice, for $q$ changing at the Fermi surface, is that
\begin{equation}
\tilde{S}_q^{(R)} = \tilde{S}_q^{(c)} \label{R=c}
\end{equation}
as given by Eqs. (\ref{SSq}) with constraints
(\ref{qconstraints1}) and with $n_{qi}$ and $\bar{n}_{qi}$ given
by Eqs. (\ref{nq3}) and (\ref{nq3a}). The effective occupation
numbers are now $n^{\tilde{q}}_{\tilde{q}i}$ (with $\tilde{q}$
defined by Eq. (\ref{hattilde})) and $\bar{n}^q_{qi}$. Note that
this time one encounters for particles a discontinuity in this
variable on the Fermi surface, i.e. for $n_{qi}=1/2$. However,
there is no such discontinuity for antiparticles.

\section{Results}
\label{sec:Results}

\subsection{Introductory remarks}
\label{sec:IR}

We first check numerically the thermodynamical consistency of
$q$-NJL.  We have the following set of differential equations:
\begin{equation}
d\varepsilon_q =T ds_q + \mu d\rho_q,\quad dP_q = s_q dT + \rho_q
d\mu, \label{def}
\end{equation}
where the energy density $\varepsilon_q = E_q/V$, the entropy
density $s_q = S_q/V$ and the density $\rho_q = N_q/V$. The
relations to be checked are
\begin{equation}
T = \frac{\partial \varepsilon_q}{\partial
s_q}\Big|_{\rho_q},~~\mu =\frac{\partial \varepsilon_q}{\partial
\rho_q}\Big|_{s_q}, ~~\rho_q = \frac{\partial P_q}{\partial
\mu}\Big|_T,~~ s_q = \frac{\partial P_q}{\partial T}\Big|_{\mu}.
\label{Tmurhos}
\end{equation}
The last two are easy to check numerically. In the first two one
has first to convert derivatives in $s$ and $\rho$ to derivatives
in $T$ and $\mu$ and to calculate
\begin{eqnarray}
T &=& \frac{\partial \varepsilon_q}{\partial s_q}\Big|_{\rho_q} =
\frac{\frac{\partial \varepsilon_q}{\partial T} + \frac{\partial
\varepsilon_q}{\partial \mu}\frac{d\mu}{d T}}{\frac{\partial
s_q}{\partial T} + \frac{\partial s_q}{\partial \mu}\frac{d \mu}{d
T}}~~{\rm where}~~\frac{d \mu}{d T} = - \frac{\frac{\partial
\rho_q}{\partial T}}{\frac{\partial \rho_q}{\partial \mu}},
\label{TTT}
\\
\mu &=& \frac{\partial \varepsilon_q}{\partial \rho_q}\Big|_{s_q}
= \frac{\frac{\partial \varepsilon_q}{\partial T} + \frac{\partial
\varepsilon_q}{\partial \mu}\frac{d\mu}{d T}}{\frac{\partial
\rho_q}{\partial T} + \frac{\partial \rho_q}{\partial \mu}\frac{d
\mu}{d T}}~~{\rm where}~~\frac{d \mu}{d T} = -
\frac{\frac{\partial s_q}{\partial T}}{\frac{\partial
s_q}{\partial \mu}}. \label{MuMuMu}
\end{eqnarray}
We have checked these relations taking as input $\mu = 322$ MeV
and $ T = 60$ MeV, for case $(a)$ with $q = 1.1$, case $(b)$ with
$q = 0.9$ and case $(c)$ with $q = 1.1$, and also for the
extensive case $q = 1$. We have also checked that these relations
are satisfied even for a broader range of values of the parameter
$q$, ranging from $q = 0.8$ to $q = 1.2$. They are all satisfied
with an accuracy of better than $1\%$.

Our previous formulation of the $q$-NJL model \cite{RWJPG}
followed essentially choice $(c)$ with regard to calculations of
the nonextensive occupation numbers $n_q$ and $\bar{n}_q$, but
their effective values, $n_q^q$ and $\bar{n}^q_q$, were wrongly
used with a constant value of the exponent $q$. The three
formulations of the $q$-NJL model presented in this work, with
$\tilde{S}^{(R=a,b,c)}_q$ defined by Eqs. (\ref{R=a}), (\ref{R=b})
and (\ref{R=c}) and described in detail in, respectively, Sections
(\ref{sec:a}), (\ref{sec:b}) and (\ref{sec:c}), were never
compared with each other. In what follows we provide such a
comparison, calculating the dependencies on chemical potential
$\mu$ and temperature $T$ of the entropy $S_q$, pressure $P_q$,
normalized density $\rho/\rho_0$, heat capacity $C_{\mu}$ and
barionic susceptibility $\chi_B$.

\subsection{Results for entropies $\tilde{S}^{(R=a,b,c)}_q$}
\label{secRfE}

\begin{figure*}[t]
\resizebox{0.5\textwidth}{!}{%
  \includegraphics{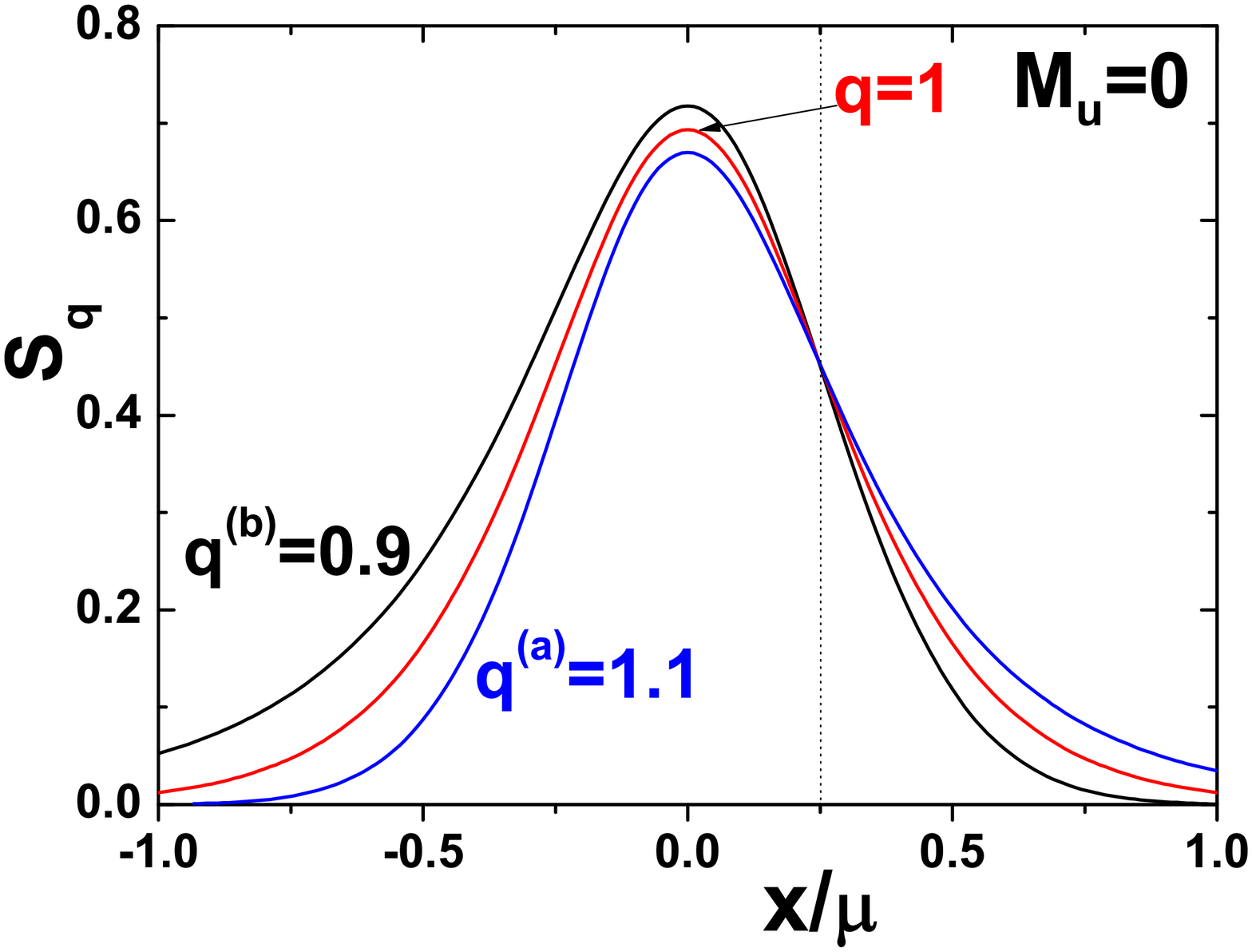}
}\hspace{5mm}
\resizebox{0.5\textwidth}{!}{%
  \includegraphics{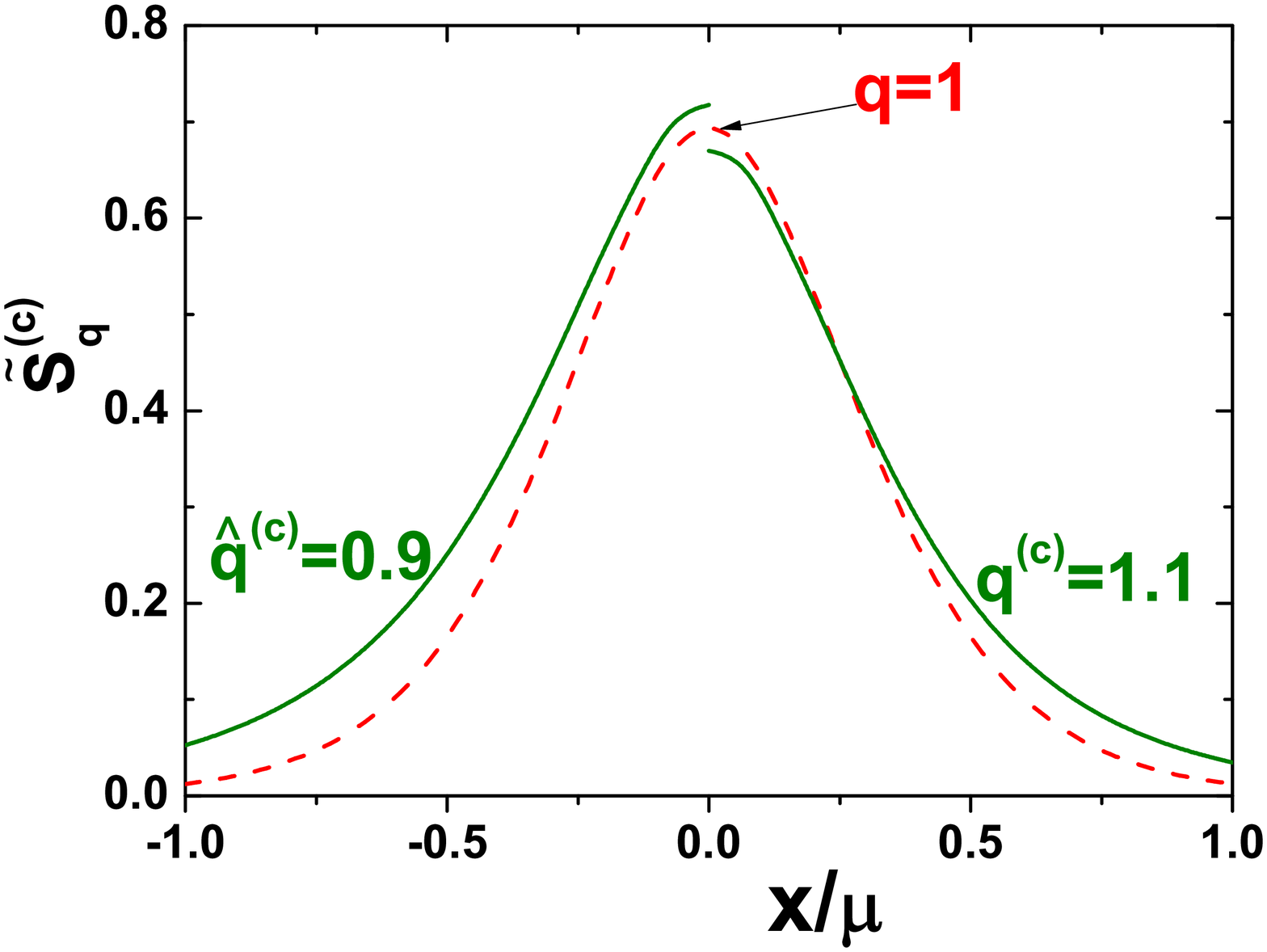}
} \vspace{-4mm}\caption{(Color online) Schematic view of $S_q =
\tilde{S}^{(R=a,b)}_q$ defined by Eqs. (\ref{R=a}) and (\ref{R=b})
(left panel) and $S_q^{(c)} = \tilde{S}^{(R=c)}$ defined  by Eq.
(\ref{R=c}) (right panel), all presented as a function of the
scaled variable $x/\mu$. The respective values of $q$ used are
shown. All these results are compared with the extensive case of
$q=1$ (blue curves). Calculations were performed assuming $M_{qi}
= 0$, $T = 60$ MeV and $\mu = 322$ MeV.  The meaning of $q$ and
$\hat{q}$ on the right panel corresponds to definition of
$\tilde{q}$ presented in Eq. (\ref{hattilde}). }
\label{S_explanation}
\end{figure*}

\begin{figure*}
\resizebox{0.5\textwidth}{!}{%
  \includegraphics{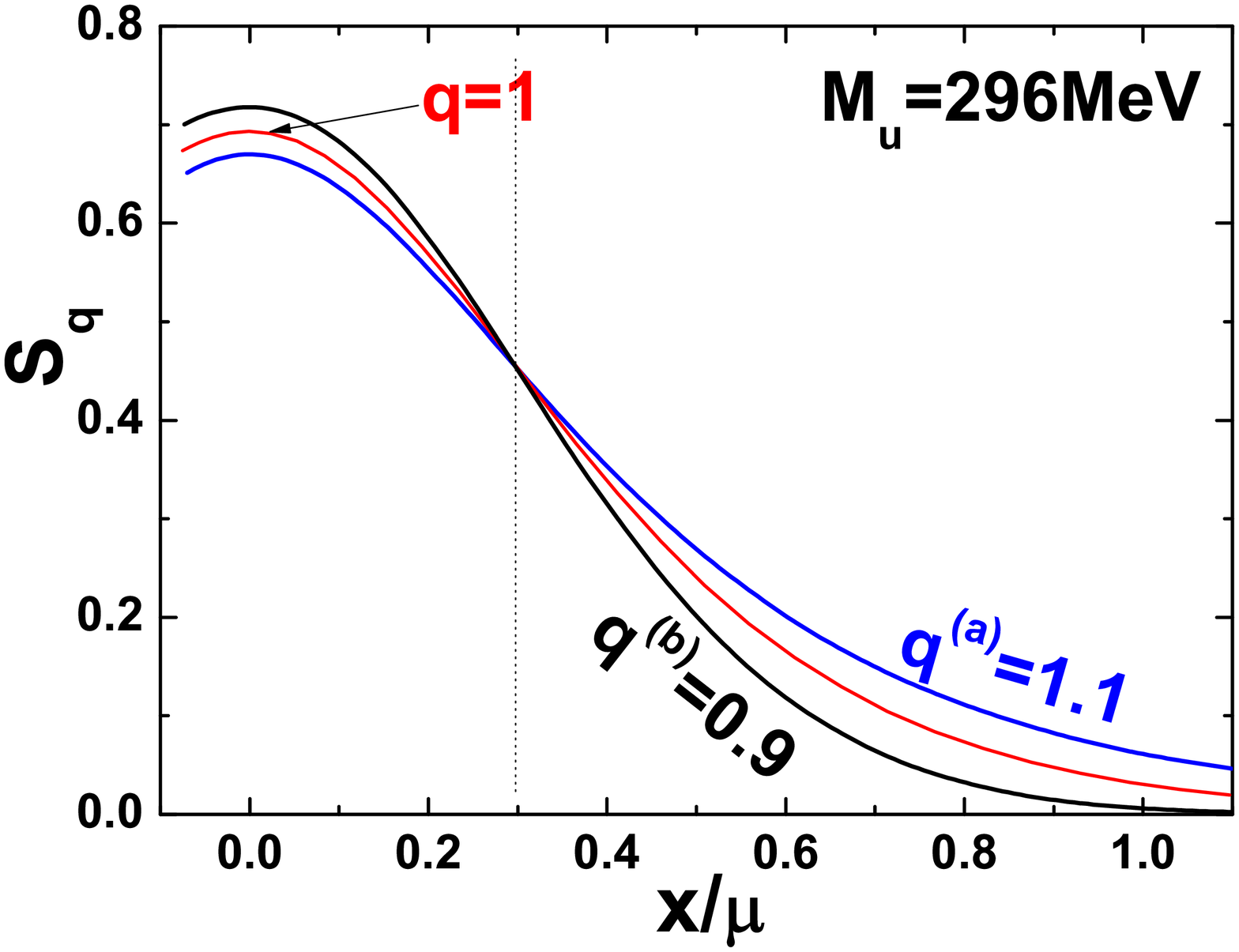}
}\hspace{5mm}
\resizebox{0.5\textwidth}{!}{%
  \includegraphics{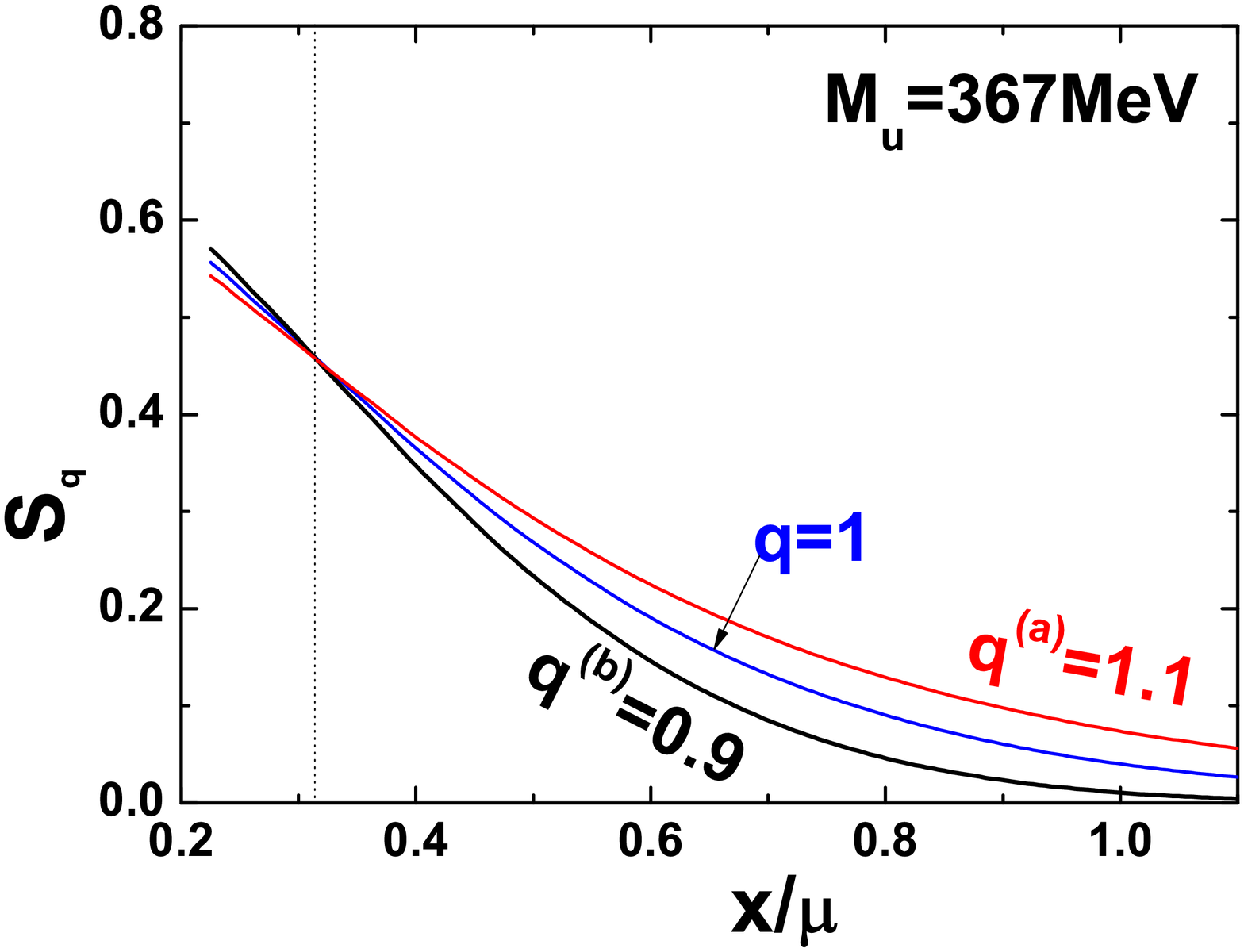}
} \vspace{-4mm}\caption{(Color online) The same as in the left
panel of Fig. \ref{S_explanation} but for different masses $M_u$.}
\label{S_explanation_1}
\end{figure*}

We start with a presentation of the entropies $\tilde{S}_q^{(R)}$
corresponding to different ways ($R= a, b, c$) of introducing
nonextensivity into the $q$-NJL model. They are shown as functions
of the scaled variable $xT/\mu$ in Figs. \ref{S_explanation} and
\ref{S_explanation_1}. The results shown will allow for a better
understanding of our further results in Section \ref{sec:SPR}. In
Fig. \ref{S_explanation} we present a schematic view of
$\tilde{S}^{(R=a,b,c)}_q$ defined by Eqs. (\ref{R=a}), (\ref{R=b})
and (\ref{R=c}). The respective values of $q$ used are $q^{(a)} =
1.1$ and $q^{(b)} = 0.9$ in the upper panel and $q^{(c)} = 1.1$ in
the lower panel. All these results are compared with the extensive
case of $q=1$. Calculations were performed assuming a
$q$-independent value of quark mass $M_u = 0$ MeV, temperature $T
= 60$ MeV and chemical potential $\mu = 322$ MeV. Note that
$\tilde{S}_q^{(c)}$, shown in the lower panel for $q^{(c)} = 1.1$,
follows for $x < 0$ the curve $q^{(b)} = 0.9$ from the upper
panel, drops down below the extensive case of $q = 1$ at $x = 0$
(on the Fermi surface), and follows for $x > 0$ the results for
$q^{(a)} = 1.1$ (crossing the curve for $q = 1$ at the same point
as in the upper panel). Fig. \ref{S_explanation_1} presents the
same entropies as in the upper panel of Fig. \ref{S_explanation}
but for (again $q$-independent) quark masses $M_u = 296$ MeV
(upper panel) and $M_u = 367$ MeV (lower panel, this is the
maximum mass of the $u$-quark which can be obtained in this
approach). Entropies $\tilde{S}_q^{(R)}$ multiplied by square
momentum, $p^2$, and with running mass $M_{qi}$, will be further
integrated and result in entropy $S_q$ in Eq. (\ref{qEnt}), which
will be shown below in Fig. \ref{S}.

Some comments are in order here. It is known that in the case when
occupation numbers do not depend on $q$ one observes the ordering
of entropies from  subextensive for $q > 1$ to superextensive for
$q < 1$ with $\tilde{S}_{q > 1} < \tilde{S}_{q = 1} < \tilde{S}_{q
<1}$ \cite{T,T1,T2,T3}. However, it turns out that in the case where
occupation numbers depend on $q$, like in our $q$-NJL model, the
situation is more complicated. Two features can be observed.
First, this ordering changes with $x$. It starts with
$\tilde{S}_q^{(a)} < \tilde{S}_{q=1} < \tilde{S}_q^{(b)}$, but
somewhere above the Fermi surface it reaches the point (marked by
the vertical line) where all these entropies coincide and, from
there, their order is reversed. For $M_u = 0$ this point is
located at $x = 0.25$, but as seen in Fig. \ref{S_explanation_1}
it shifts slowly to higher values of $x$ with increasing mass
$M_{qi}$. At the same time the minimum value of $x$ (corresponding
to $p=0$ and equal to $\left[ \left( M^2_u + p^2\right)^{1/2} -
\mu\right]/T$, cf.  Eq. (\ref{qxbarx})) shifts with increasing
$M_u$ towards higher values, nearer the point where all entropies
coincide and their order is reversed. This effect is strongly
enhanced for higher momenta and, eventually, almost all
contributions to the final entropy will come from this region.
This will result in the ordering of entropies seen below in Fig.
\ref{S}. The same shift would be observed for case $(c)$ (not
shown here since, as was stated previously, the results for this
case can be obtained by a suitable composition of the results for
cases $(a)$ and $(b)$). This means that for higher masses and
momenta the dominant component will therefore be given by $q^{(a)}
= 1.1$ (which will show up in Fig. \ref{S} below).

We close this section by noting that it can be checked that such
behavior of the entropies $\tilde{S}_q^{(R)}$ is caused by
differences in the tails of the occupation number distributions
$n_q(x)$ as functions of $x$, additionally enhanced by the fact
that in the definition of different $\tilde{S}_q^{(R)}$ they enter
as powers of $q$, $n_q^q(x)$.  For $x=0$ all distributions
coincide to $n_q(0)= 1/2$. However, in case $(c)$, where at $x=0$
the power index $q$ changes to $2-q$ (for example from  $q=1.1$ to
$2-q = 0.9$) $n_q^q(0) \neq n_q^{(2-q)}(0)$. All these results
were calculated neglecting the contribution of the sea (i.e.
neglecting antiparticles). We have checked that their contribution
does not change the conclusions reached here, their effect being
of the order of $~1\%$ only.
\begin{figure*}
\resizebox{0.5\textwidth}{!}{%
  \includegraphics{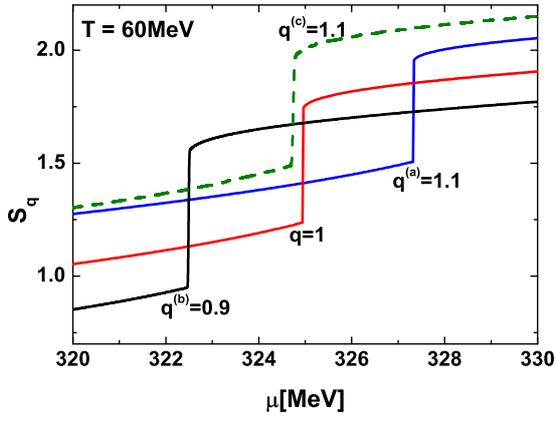}
}\hspace{5mm}
\resizebox{0.5\textwidth}{!}{%
  \includegraphics{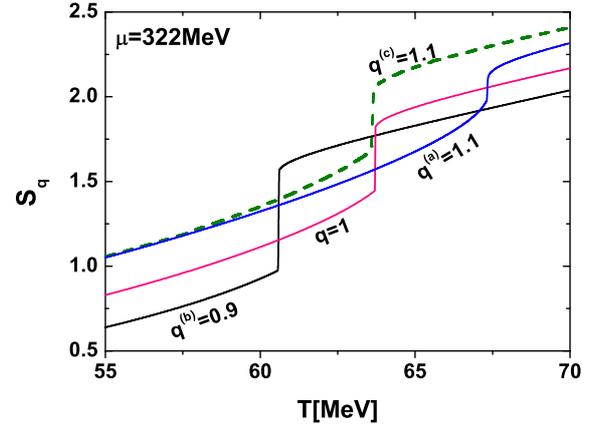}
} \vspace{-4mm}\caption{(Color online) Entropy as a function of
the chemical potential (left panel) and temperature (right panel)
calculated for different values of $q$ corresponding to different
realizations of the $q$-NJL model and compared with the BG case of
$q=1$.} \label{S}
\end{figure*}

\begin{figure*}
\resizebox{0.5\textwidth}{!}{%
  \includegraphics{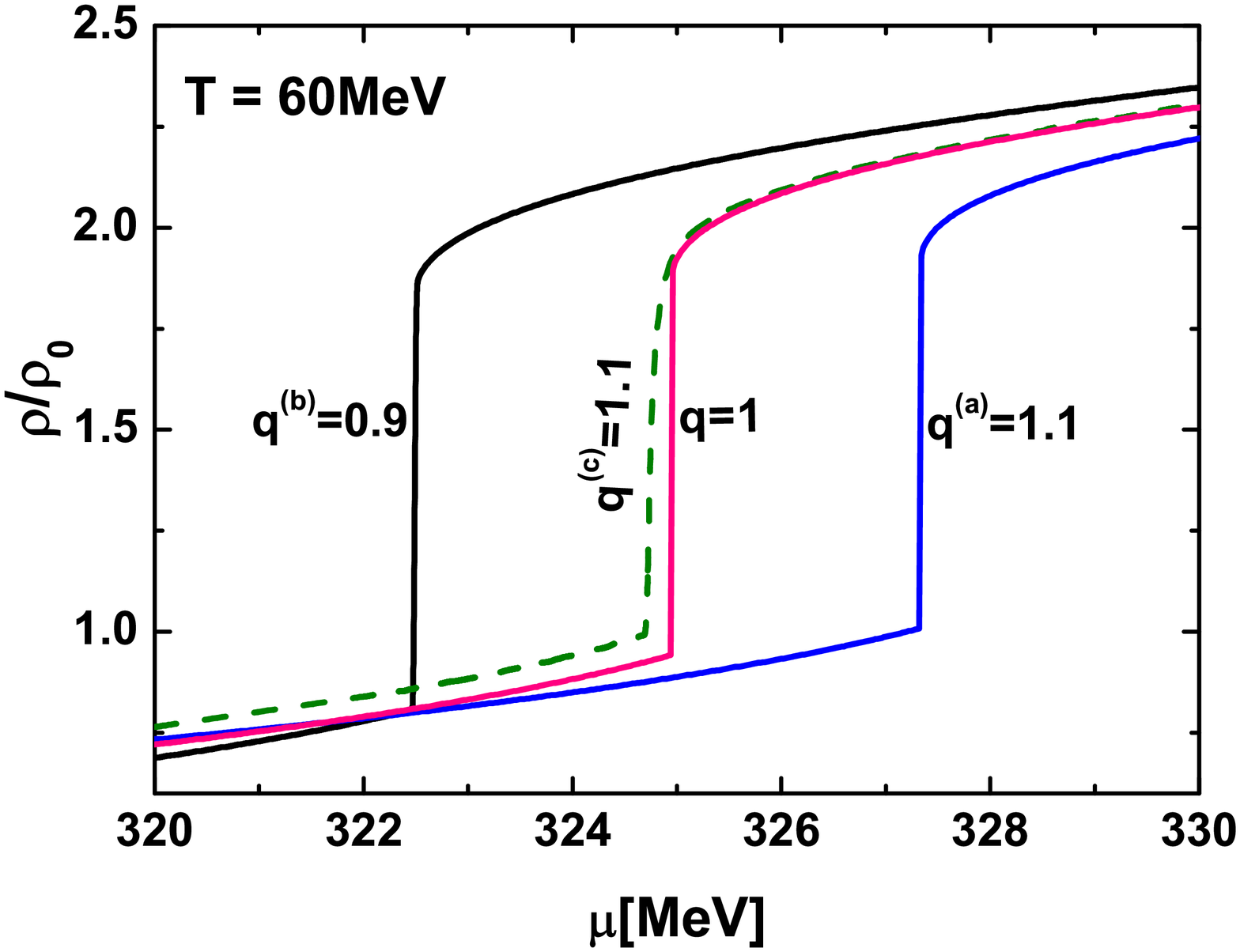}
}\hspace{5mm}
\resizebox{0.5\textwidth}{!}{%
  \includegraphics{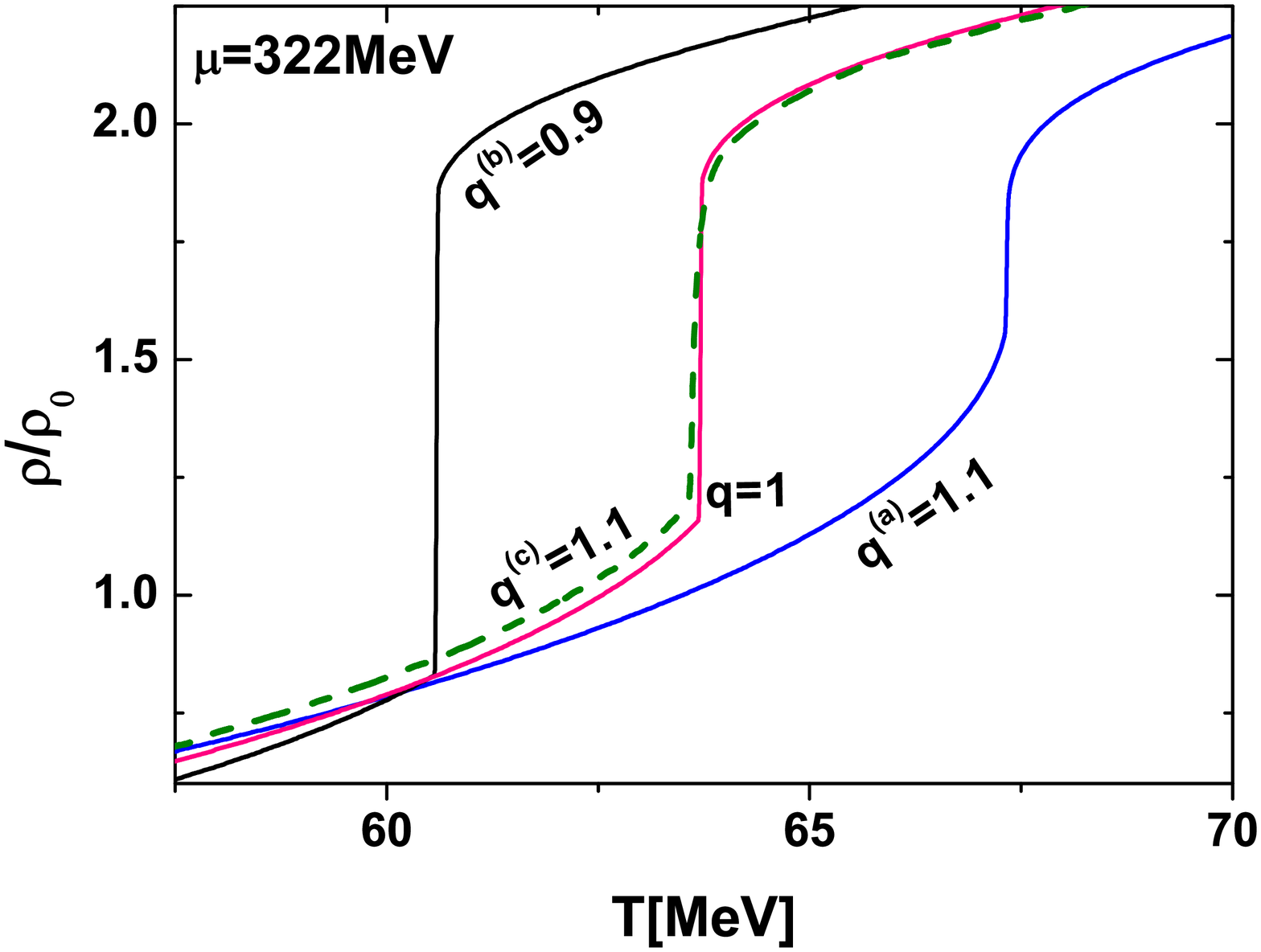}
} \vspace{-4mm}\caption{(Color online)  Compression $\rho/\rho_0$
as a function of the chemical potential (left panel) and the
pressure (right panel) for different values of $q$ corresponding
to different realizations of the $q$-NJL model and compared with
the BG case of $q=1$.} \label{R}
\end{figure*}

\subsection{Results for entropy, pressure and compression}
\label{sec:SPR}

\begin{figure*}
\resizebox{0.5\textwidth}{!}{%
  \includegraphics{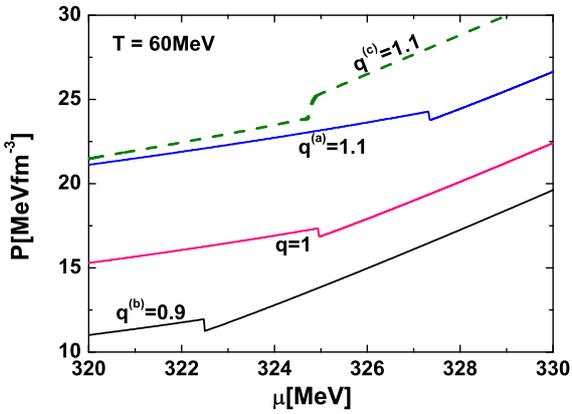}
}\hspace{5mm}
\resizebox{0.5\textwidth}{!}{%
  \includegraphics{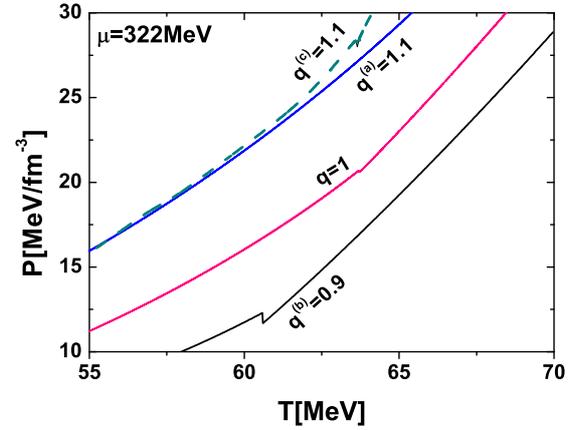}
} \vspace{-4mm}\caption{(Color online) Pressure as a function of
the chemical potential (left panel) and pressure (right panel) for
different values of $q$ corresponding to different realizations of
the $q$-NJL model and compared with the BG case of $q=1$.}
\label{P}
\end{figure*}

After the preparatory explanations in the previous section we
present now results for entropy $S$ (Fig. \ref{S}), pressure $P$
(Fig. \ref{P}) and density $\rho$ (Fig. \ref{R}).  Calculations
were performed for $T = 60$ MeV and $\mu = 322$ MeV, for different
values of $q$ corresponding to different realizations of the
$q$-NJL model and compared with the BG case of $q=1$. This choice
of values of $T$ and $\mu$ is dictated by the fact that, as can be
seen in Table \ref{Table_1}, they are near to the critical values
corresponding to the values of nonextensive parameter $q$
considered here. For this choice the critical effects displayed in
the presented figures are most visible. Chemical potentials and
temperatures at which one observes rapid changes of entropy (Fig.
\ref{S}) and small jumps in the pressure (Fig. \ref{P}),
correspond to regions of phase transition in which the density of
fermions changes rapidly, approximately from $\rho_0$ to
$2\rho_0$, as seen in Fig. \ref{R}.

The general characteristic feature of these results is the
horizontal shift of the phase transition point and of all curves
for entropy $S$ and density $\rho/\rho_0$, when calculated for a
fixed value of temperature $T = 60$ MeV, towards smaller values of
$\mu$ for $q < 1 $ (case $(b)$) and towards greater values of
$\mu$  for $q^{(a)} = 1.1$ (case$(a)$). For densities below and
above well separated critical points all curves converge. At the
same time $S_{q=1.1}^{(a)} > S_{q=1}$, i.e. it exceeds the BG
entropy, whereas $S^{(b)}_{q=0.9}$ is smaller than the BG entropy.
In case $(c)$, in which $q = 1.1$ above the Fermi surface and
changes to $\hat{q} = 2 - q = 0.9$ below it, one observes more
complicated behavior. Whereas the transition point remains nearly
the same as in the BG case, the value of entropy $S_q^{(c)}$
exceeds considerably that of $S^{(a)}_{q=1.1}$ above the phase
transition point and practically coincides with it below this
point. This behavior reflects that of $\tilde{S}_q^{(R=a,b,c)}$
shown in Figs. \ref{S_explanation} and \ref{S_explanation_1} (one
must remember that $M_u$ is a minimum above the phase transition
point and a maximum below it)\footnote{ Similar increase of
entropy corresponding to our choice $(c)$ was also observed in
hadrodynamical description of the nuclear matter
\cite{AD_stars}.}.

Behavior of pressure $P$ is shown in Fig. \ref{P}. Its increase as
a function of $\mu$ is dictated by the derivative $\rho =
\partial P/\partial \mu|_T$  (and depends on the behavior of
$\rho$ presented in Figure \ref{R}), whereas its increase as a
function of $T$ is given by the derivative $s =\partial P/\partial
T |_{\mu}$ (and is given by the behavior of $S$ in Fig. \ref{S}).
Looking at Figs. \ref{S}-\ref{R} from this perspective, we observe
generally that both the entropy and density are lower below the
phase transition than after it.  The corresponding increase of
pressure before the phase transition is therefore weaker than in
unconfined phase. This behavior is more pronounced when considered
as a function of temperature $T$ rather than as a function of
chemical potential $\mu$. Note that because, as seen in Fig.
\ref{R}, densities are very similar for different realizations of
the $q$-NJL model, the resulting pressure curves as a function of
chemical potential are almost parallel. Because entropies differ
much more strongly the dependence on temperature is more divergent
for different realizations of the $q$-NJL model.

\subsection{In the vicinity of the phase transition point}
\label{sec:CmuChiPh}

We shall present now the behavior in the vicinity of critical
values of $T$ and $\mu$ and present results for pressure,
compression, heat capacity $C_{\mu}$, baryon number susceptibility
$\chi$ and the phase diagram.
\begin{figure*}
\resizebox{0.5\textwidth}{!}{%
  \includegraphics{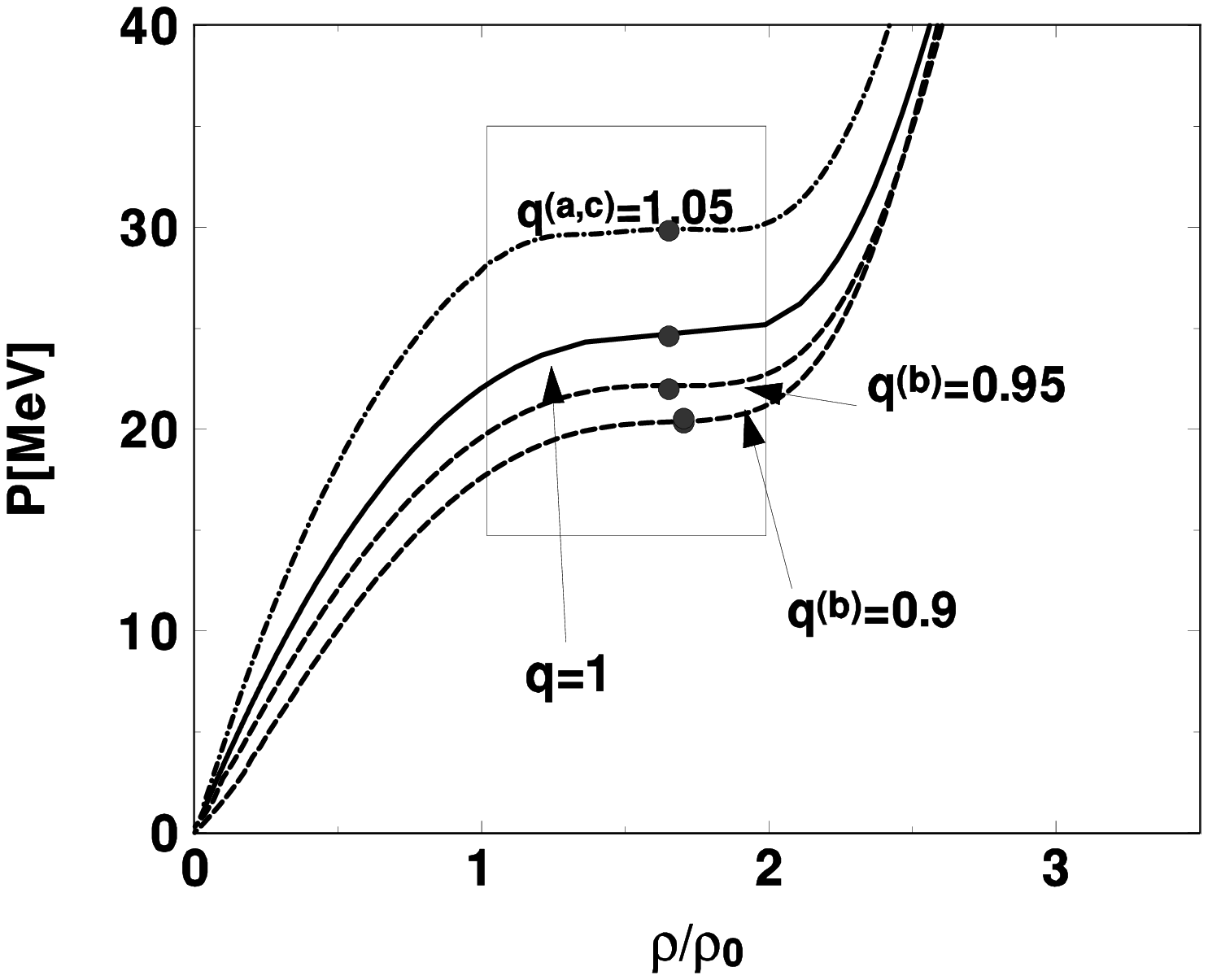}
}\hspace{5mm}
\resizebox{0.5\textwidth}{!}{%
  \includegraphics{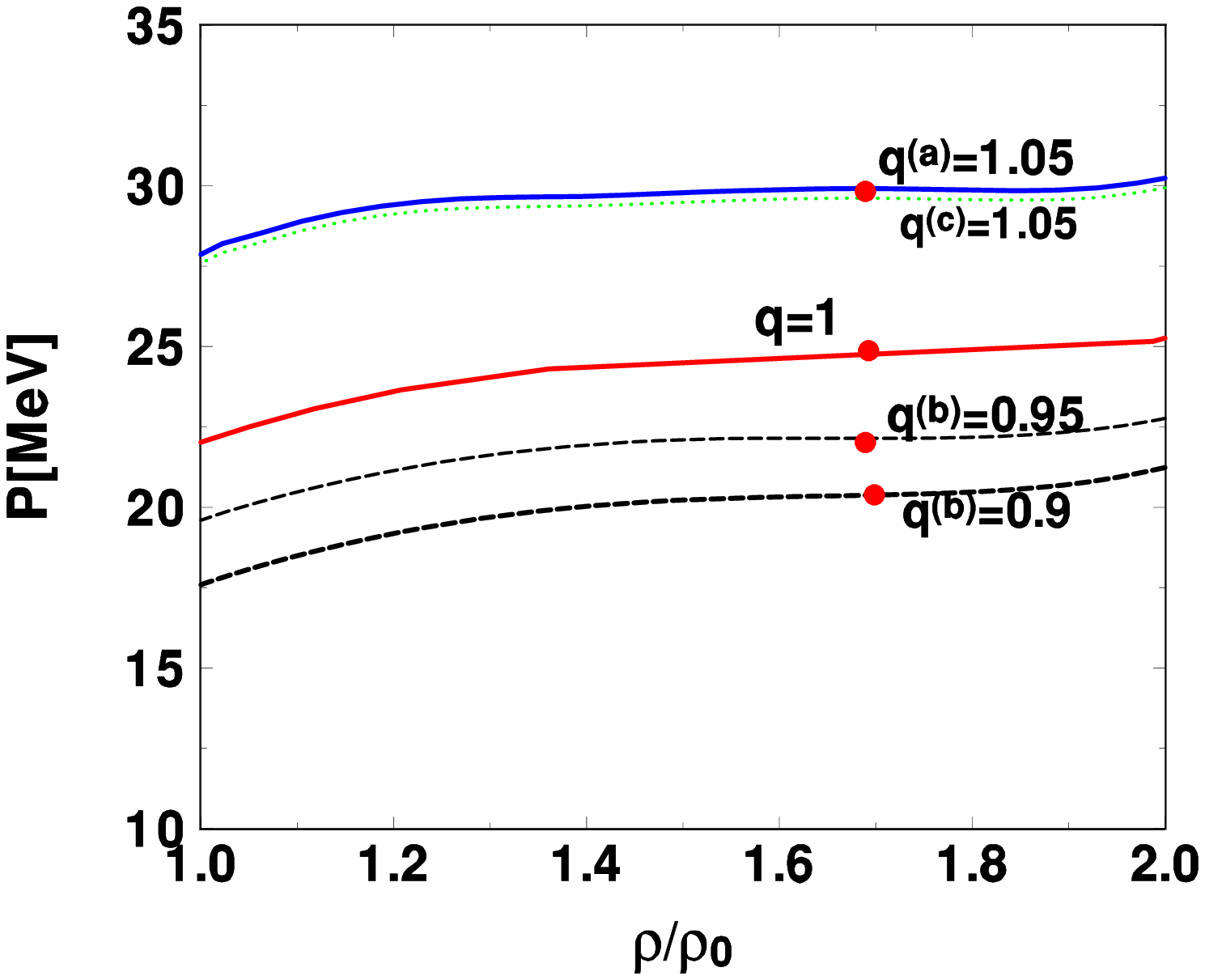}
} \vspace{-4mm}\caption{(Color online)  The pressure at critical
temperature $T_{cr}$ as a function of compression $\rho/\rho_0$
calculated for different values of $q$ corresponding to different
realizations of the $q$-NJL model and compared with the BG case of
$q=1$ (the area marked on the upper panel is shown in detail in
the lower panel). The dots indicate the positions of the
inflection points for which the first derivative of pressure by
compression vanishes.} \label{P_at_Tc}
\end{figure*}
Fig. \ref{P_at_Tc} presents the pressure at the critical
temperature $T_{cr}$ (critical isotherms) for the corresponding
critical values of the chemical potential, $\mu_{cr}$, as a
function of compression $\rho/\rho_0$. The values of $T_{cr}$ and
$\mu_{cr}$ determine the critical values of pressure and density
corresponding to the inflection points for which the first
derivative of pressure by compression vanishes. They are listed in
Table \ref{Table_1}. Remarkably, in all these cases the critical
density remains practically the same. The other interesting
observation is that the results for choice $(a)$ essentially
coincide with those for choice $(c)$. For $q > 1$, we observe an
increase of the pressure $P$ (corresponding to a similar increase
of the entropy in this case observed in Fig. \ref{S}). This
increase may simulate some additional repulsion between massive
quarks constituting nucleons at short distances (which is present
in quark-meson coupling models based on the mean field approach
\cite{QMC,QMC1}). On the contrary, for $q < 1$ we observe a decrease of
the critical pressure (corresponding to a decrease of entropy,
seen in Fig. \ref{S}). This could simulate confinement of quarks
in nucleons which introduces restrictions in phase space by
changing their Fermi motion.
\begin{figure*}
\resizebox{0.5\textwidth}{!}{%
  \includegraphics{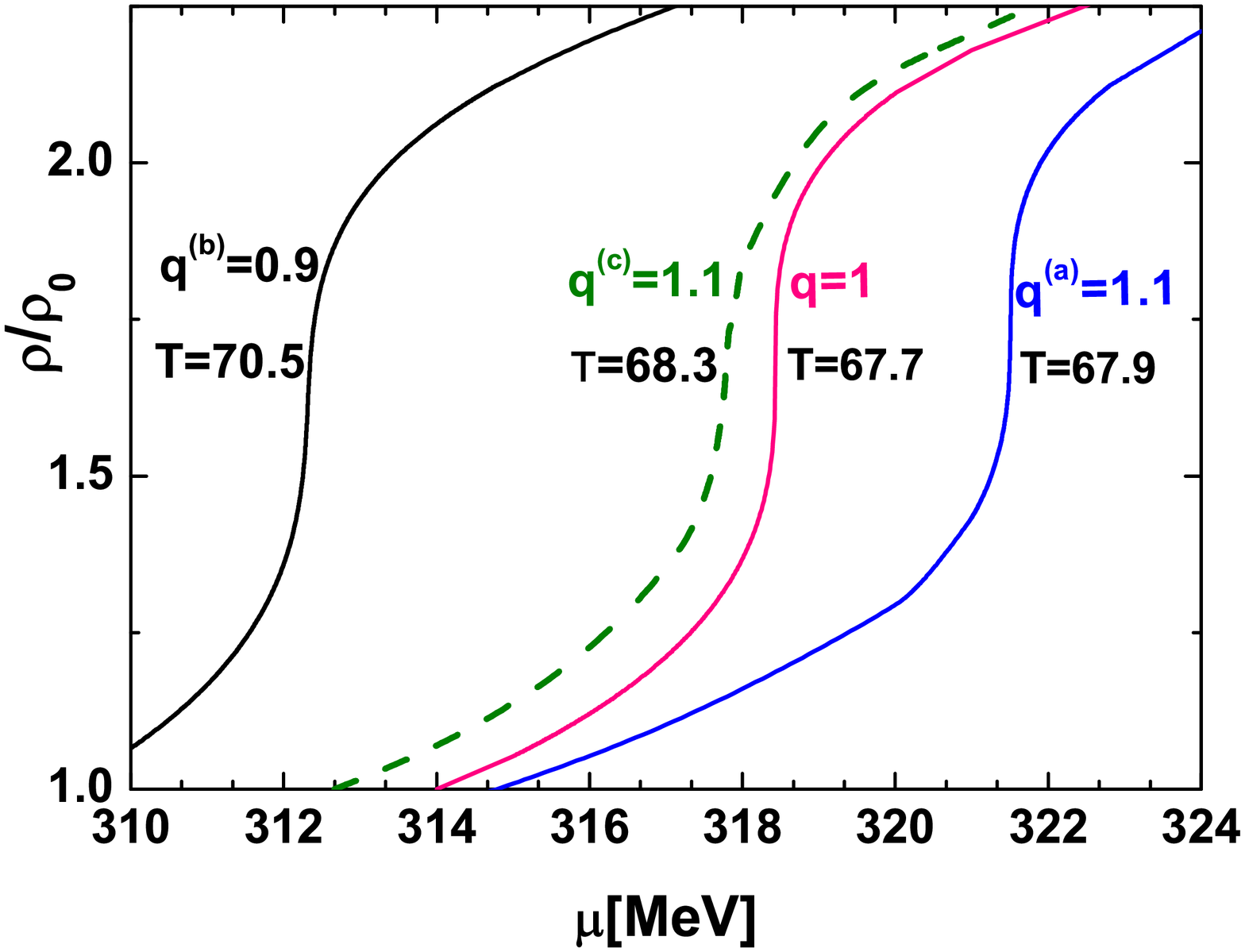}
}\hspace{5mm}
\resizebox{0.5\textwidth}{!}{%
  \includegraphics{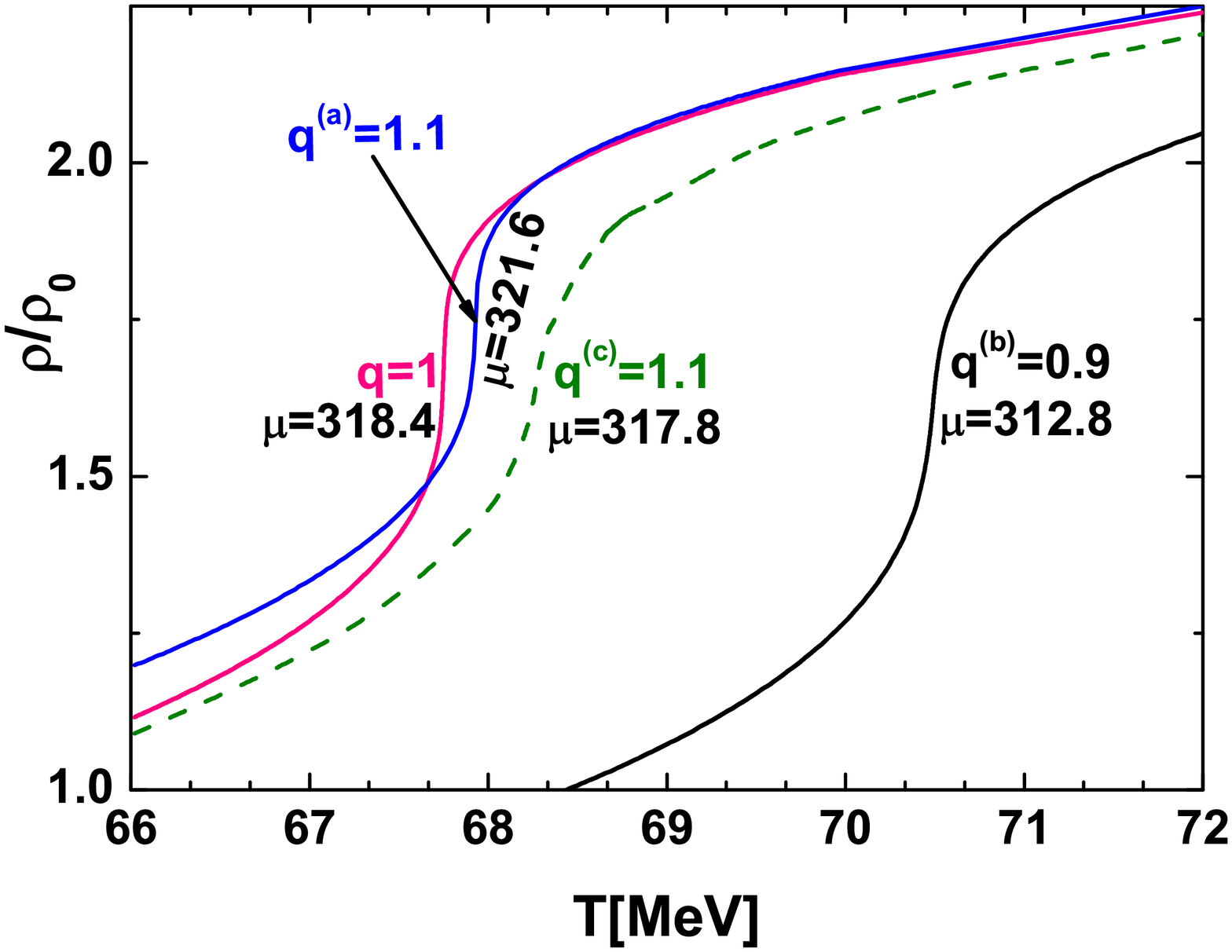}
} \vspace{-4mm}\caption{(Color online)  Compression $\rho/\rho_0$
as a function of chemical potential (left panel) and temperature
(right panel) calculated in the vicinity of the phase transition
for different values of $q$ corresponding to different
realizations of the $q$-NJL model and compared with the BG case of
$q=1$.} \label{R1}
\end{figure*}

\begin{table}[h]
\caption { The values of the critical temperature, $T_{cr}$, and
the critical chemical potential, $\mu_{cr}$, for different
realizations of the $q$-NJL model.} \vspace*{0.2cm}
\begin{center}
\begin{tabular}{|l|c|c|}
\hline
           &~ $T_{cr}$ [MeV]~ &~ $\mu_{cr}$ [MeV]~\\
\hline
$q^{(a)} = 1.1$  &  $67.9$          &  $321.6$ \\
$q^{(b)} = 0.9$  &  $70.5$          &  $312.3$ \\
$q^{(c)} = 1.1$  &  $68.3$          &  $317.8$ \\
$q = 1$    &  $67.7$          &  $318.4$ \\
\hline
\end{tabular}
\end{center}
\label{Table_1}
\end{table}
In Fig. \ref{R1} we present compression as a function of chemical
potential $\mu$  and  temperature $T$ but now, unlike in Fig.
\ref{R}, calculated in the vicinity of the phase transition point
(the corresponding values of $T_{cr}$ and $\mu_{cr}$ are given in
Table \ref{Table_1}). Because both $T$ and $\mu$ now change their
values, the observed behavior of $\rho/\rho_0$ is very different
from the previously one. It is dictated by the fact that, as seen
in Table \ref{Table_1}, in our $q$-NJL model we have
$T_{cr}^{(q=1)} < T_{cr}^{(a)} < T^{(c)}_{cr} < T^{(b)}_{cr}$ and
$\mu^{(a)}_{cr} > \mu^{(q=1)}_{cr} > \mu^{(c)}_{cr} >
\mu^{(b)}_{cr}$.

\begin{figure*}
\resizebox{0.5\textwidth}{!}{%
  \includegraphics{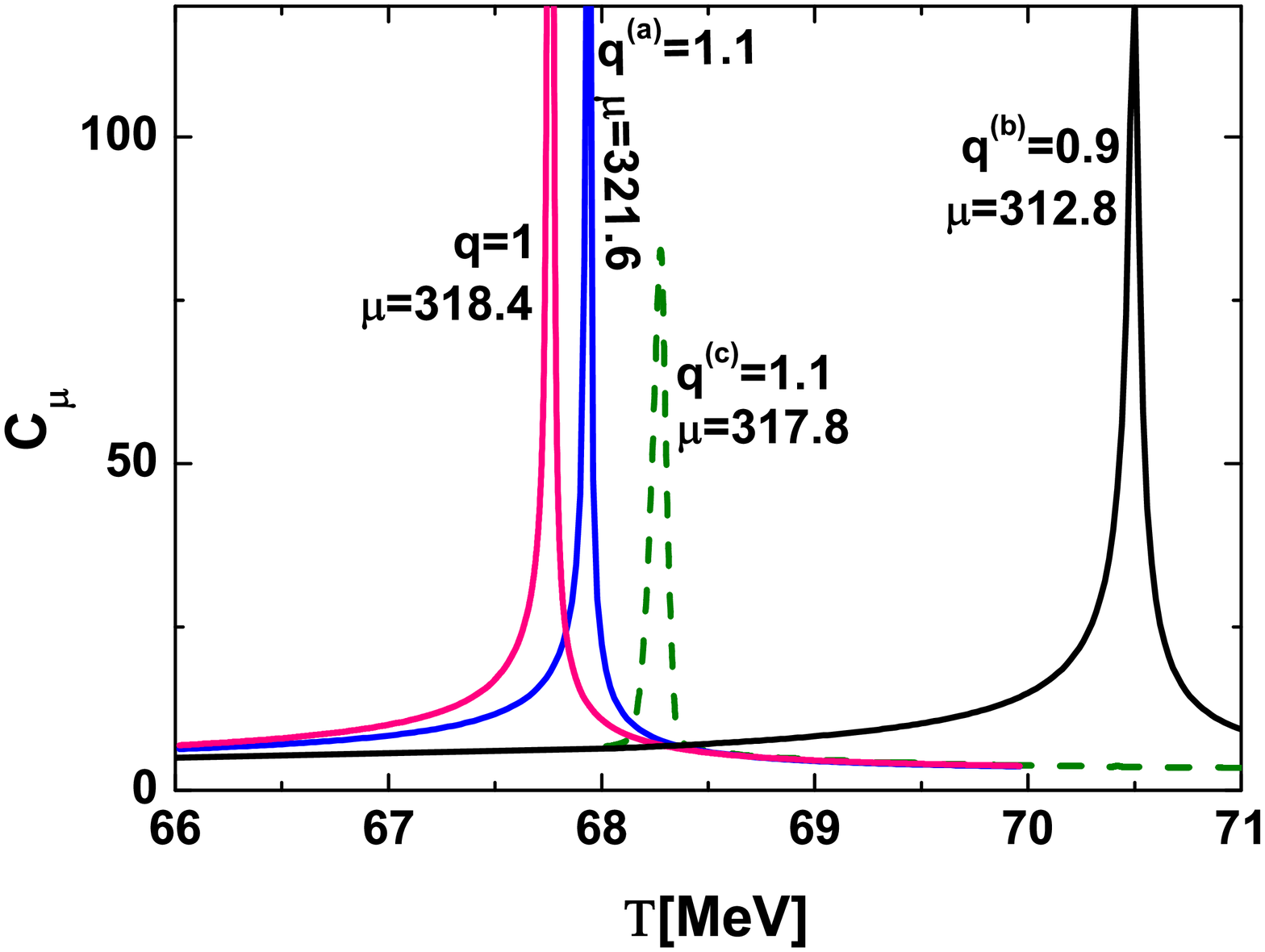}
}\hspace{5mm}
\resizebox{0.5\textwidth}{!}{%
  \includegraphics{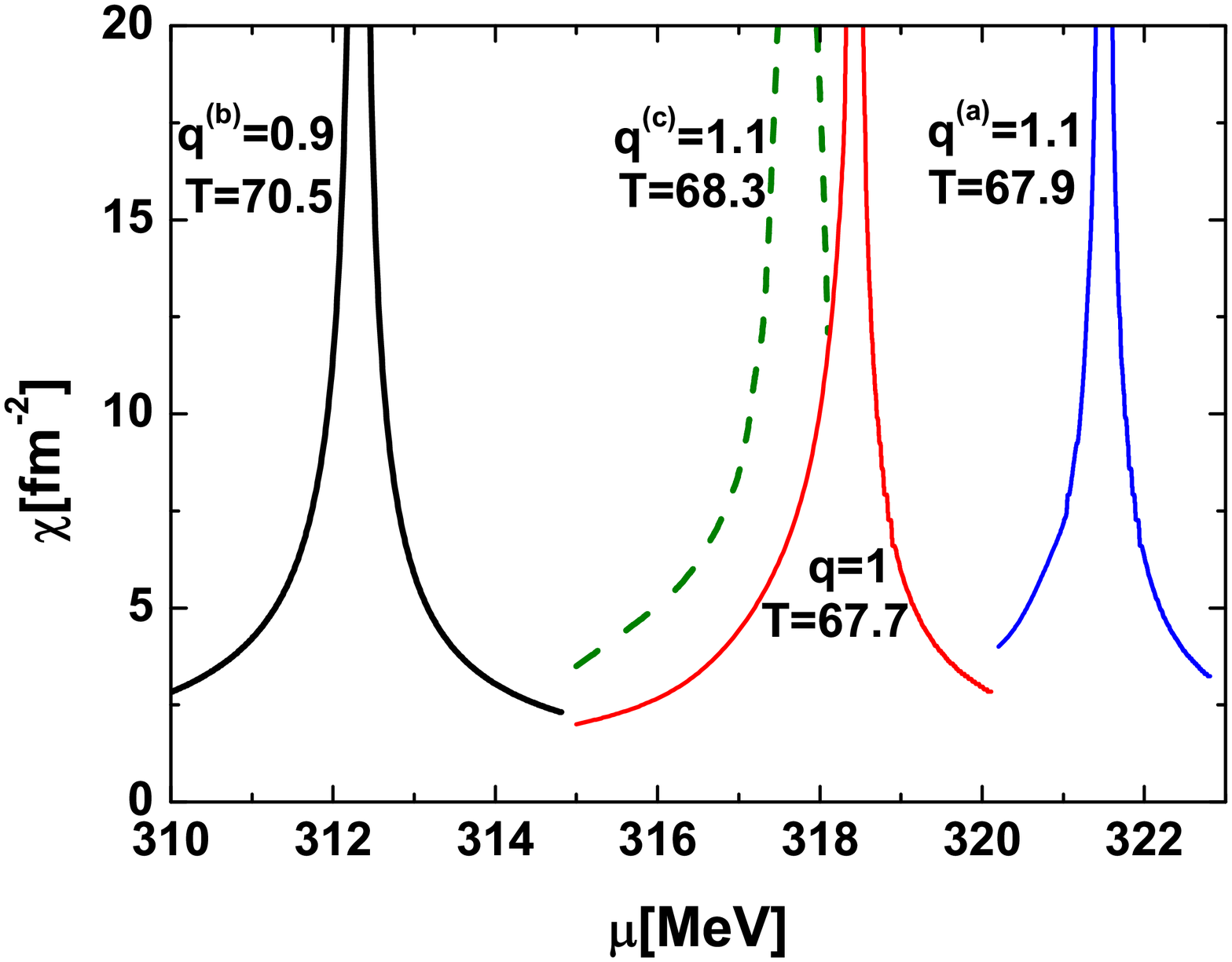}
} \vspace{-4mm}\caption{(Color online) Left panel: Heat capacity
as a function of temperature calculated in the vicinity of the
phase transition point for different values of $q$ corresponding
to different realizations of the $q$-NJL model and compared with
the BG case of $q=1$. The respective values of chemical potential
used are indicated. Right panel: Susceptibility as a function of
chemical potential  calculated in the vicinity of the phase
transition point, and for different values of $q$ corresponding to
different realizations of the $q$-NJL model and compared with the
BG case of $q=1$. The respective values of the temperatures are
indicated. } \label{C_sus}
\end{figure*}
\begin{figure*}
\resizebox{0.5\textwidth}{!}{%
  \includegraphics{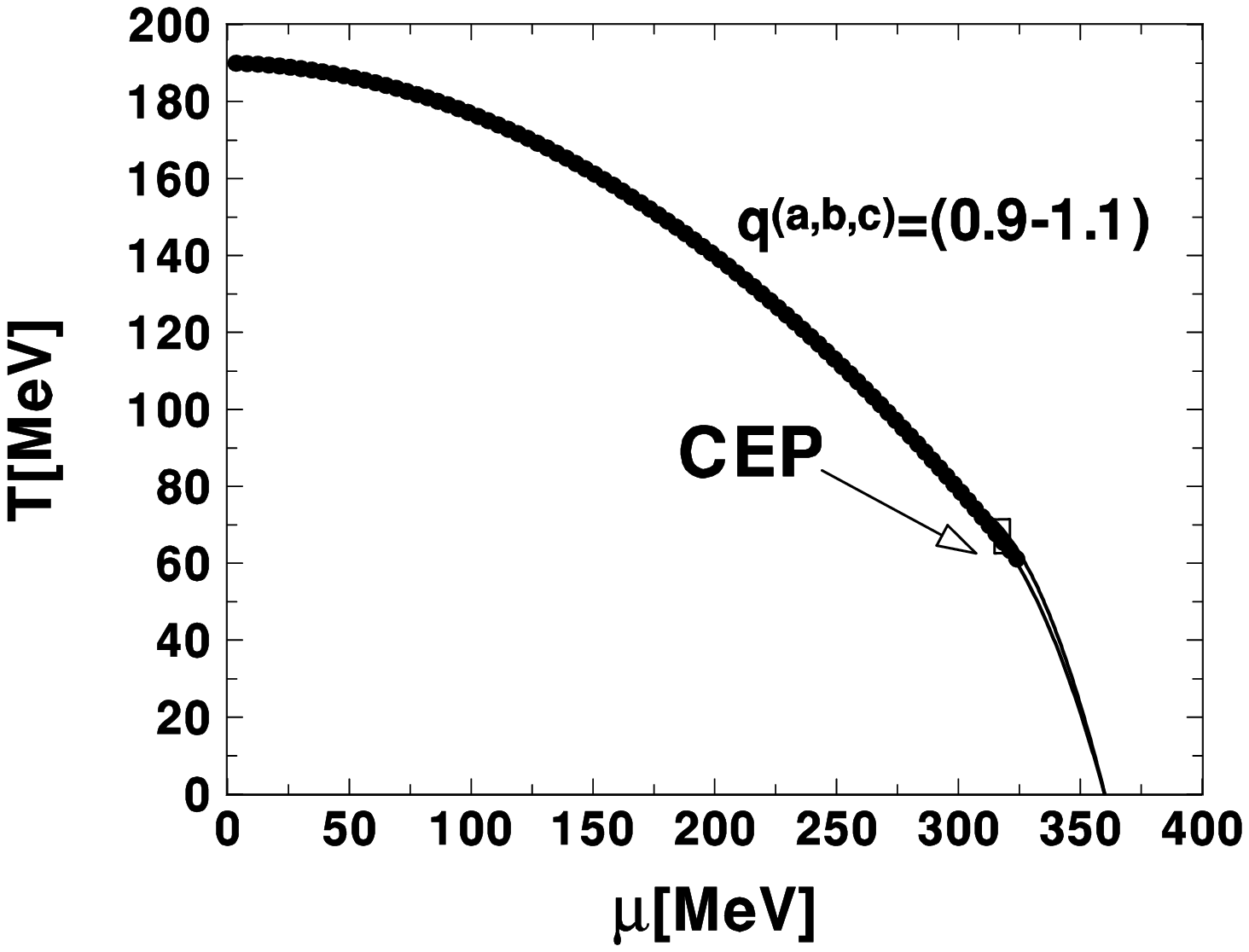}
}
\resizebox{0.55\textwidth}{!}{%
  \includegraphics{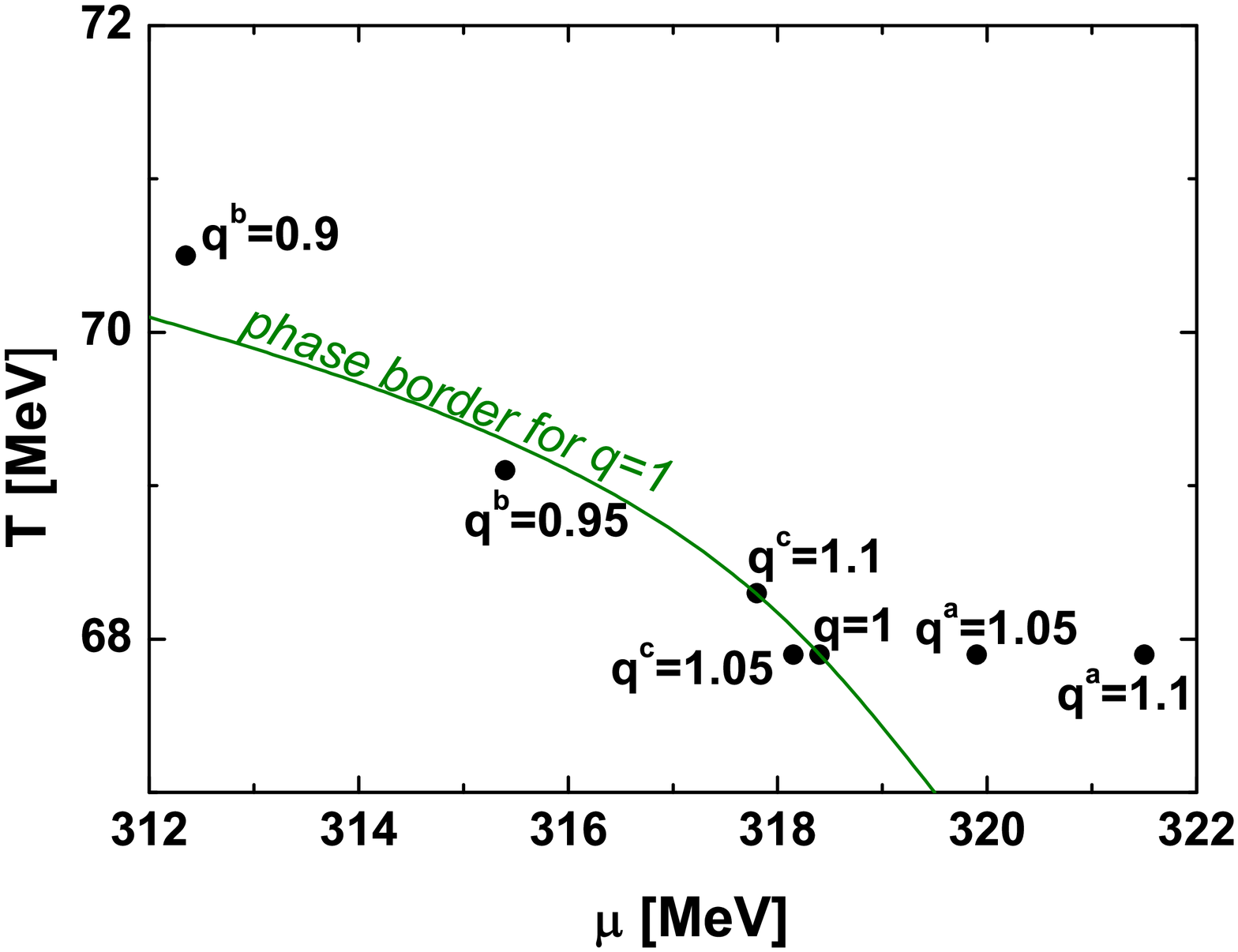}
} \vspace{-4mm}\caption{(Color online)  Phase diagram in the
$q$-NJL model in the $T-\mu$ plane for different values of $q$
corresponding to different realizations of the $q$-NJL model and
compared with the BG case of $q=1$. The left panel shows a general
view where, for the scale used, all curves essentially coincide.
The right panel shows an enlarged region near the critical point
(CEP).} \label{Phases}
\end{figure*}
In Fig. \ref{C_sus} we present our results for, respectively, heat
capacity (not calculated before in \cite{RWJPG,RWConf,RWConf1}) and baryon
susceptibilities. Both were calculated for the critical values of
the chemical potential and temperature listed in Table
\ref{Table_1}. As one can see in left panel of Fig. \ref{C_sus},
for $q^{(b)} = 0.9$ the critical point is shifted towards higher
values of temperature, $T_{crit} = 70.5$ MeV. In this region we
also observe that the chemical potential at the critical point is
lower, $\mu_{crit} = 312$ MeV, in comparison to its extensive
value for $q = 1$, $\mu_{crit}=318.4$ MeV. As a result, the
critical points for $q^{(b)} < 1$ lie very close the extensive
phase border, as seen in Fig. \ref{Phases}. For $q^{(a)} = 1.1$
(and partially also for $q^{(c)} = 1.1$) the critical temperature
$T_{crit}$ remains very near its extensive value for $q = 1$,
$T_{crit} = 70.5$ MeV, but the critical chemical potential is
shifted towards larger values, see Fig. \ref{Phases} and Fig.
\ref{C_sus} (right panel).

As seen in Fig. \ref{R1} and Fig. \ref{Phases} the results for
$T_{crit}$ are for choice $(c)$ are very near to those for choice
$(a)$ whereas the results for $\mu_{crit}$ are for choice $(c)$
very near to the results for $q =1$. Consequently, whereas the
critical coordinates in cases $(a)$ and $(b)$ depart from the
extensive phase border line in Fig. \ref{Phases}, the critical
coordinates for choice $(c)$ are very close to the extensive limit
of $q = 1$. This reflects the fact that in choice $(c)$ we use at
the same time both $q > 1$ and $q < 1$ (depending on the region of
phase space actually considered) minimizing therefore the
departure from the extensive dynamics.

The results presented in Figs. \ref{Phases}, \ref{P_at_Tc} and in
the right panel of \ref{C_sus} can be compared with the
corresponding results obtained in our previous version of the
$q$-NJL model (with Figs. 4 and 8 in \cite{RWJPG}, with Fig. 1 in
our first paper in \cite{RWConf,RWConf1} and with Figs. 1-3 and 5 in our
second paper there). This allows us to estimate what kind and how
big are the changes introduced by proper treatment of
thermodynamical consistency. Quantitatively the corresponding
results look similar, there are differences in what concerns
sensitivity to deviation from extensivity leading to
quantitatively similar effects: the previous $|q - 1|\sim 0.02$
should be confronted with the present $\sim 0.1$. The detailed
shape of the phase diagram or detailed shapes of some
distributions are also different, although, not dramatically so.
The most visible, but still rather small, difference concerns the
case with $q < 1$. For example, the previously presented $\chi$
was much smaller and broader than the corresponding result for $q
> 1$, at present they are comparable in shape and height (compare
right panel of our Fig.\ref{C_sus} and Fig.5 in the second paper
of \cite{RWConf,RWConf1}). Also, the sensitivity to changes in $q$ in the
vicinity of the phase transition point seems to be reversed.
Comparing our Fig. \ref{P_at_Tc} and Fig. 1 in \cite{RWConf,RWConf1} one
notices greater changes for $q < 1$ than for $q
> 1$ observed before, and the reverse behavior is seen in the present results.
This is so far the most visible difference resulting from the fact
that now $q < 1$ is described entirely by case $(b)$ only. The
most similar to what was used before in \cite{RWJPG,RWConf,RWConf1} is
case $(c)$, already explained in detail.

\section{Summary and conclusions}
\label{sec:Summary}

The results presented above confirm the previously reached
conclusions concerning the applicability and usefulness of the
nonextensive approach as one of the possible extensions of the
original mean field theory, extending considerably the range of
its applicability. However, this time they were obtained in
compliance with all the conditions that must be satisfied when
implementing a nonextensive approach to systems with particles and
antiparticles and a nonzero chemical potential (which were so far
overlooked) \cite{JR}. Out of three possible formulations
investigated in detail in sections (\ref{sec:a}), (\ref{sec:b})
and (\ref{sec:c}) (and denoted by $(a)$, $(b)$ and $(c)$,
respectively), the most consistent, in our opinion, is approach
$(a)$ for $q > 1$ or $(b)$ for $ q < 1$.

Approach $(c)$ requires more attention. There are two features
which should be commented on. The first is that our results in
Fig. \ref{S} show that the jump in entropy $S^{(c)}$ (for
$q^{(c)}=1.1$) occurs almost at the same point as that for the
extensive case of $q = 1$. This is because in case $(c)$ we have,
in fact, two values of the nonextensive parameter: $q$ above the
Fermi surface and $2-q$ below it (equal, correspondingly, to $1.1$
and $0.9$ in our case). As can be seen in Fig. \ref{S}, the
corresponding shifts for $q > 1$ and $ q < 1$ are in opposite
directions. Therefore, we see here a kind of cancellation of two
opposite effects. The second feature, visible in Fig.
\ref{S_explanation} (lower panel) is the jump in entropy on the
Fermi surface (which occurs because of the above mentioned change
in nonextensivity parameter there). However, in nuclear matter we
do not observe a sharp change of entropy on the Fermi surface
(i.e. for $x = 0$) in the extensive situation of $q = 1$ which
could possibly be modeled by introducing a nonextensive
environment with $q \neq 1$. The transition from the bound state
to the continuum is always gradual\footnote{A singularity at the
Fermi surface reported recently in \cite{FermiJump} occurs only as
a quantum effect at the vanishing temperature. On the other hand,
such behavior is observed in solid state physics, cf. for example
\cite{Solid}. In nuclear matter one could think of replacing the
jump in $q$ in approach $(c)$ from $q$ to $2-q$ by some smooth
transition taking place between, say, $x - \epsilon$ and
$x+\epsilon$. In fact, similar propositions in this direction,
using a suitable modifications of the occupation numbers for
$x<0$, were already presented and discussed in
\cite{DeppmanJ,BSZ}. This, however, requires a completely new and
demanding analysis of the nonextensive approach, which is beyond
the scope of the present work.} \cite{Pairing,Pairing1}.

As discussed before in Section \ref{sec:c}, case $(c)$ was
proposed in order to avoid Tsallis' cut-offs introduced in cases
$(a)$ and $(b)$ (cf. Eq. (\ref{limqg1p})). These cut-offs result
in the artificial fixing of the respective occupation numbers in
some regions of phase space to be equal either to unity
(particles) or zero (antiparticles) (cf. Eqs. (\ref{expqg1}) and
(\ref{limqg1p}) or  (\ref{expql1})-(\ref{limql1a})). They are
therefore regarded as unjustified additional ingredients of the
$q$-NJL model. However, it seems that the price of introducing
case $(c)$, which eliminates the need for such restrictions, is
too high to be acceptable for the reasons mentioned above. The
cut-off of part of the phase space seems to be more reasonable in
this respect. To add to these arguments, note that when crossing
the Fermi level the quark mass does not change, whereas it changes
in the phase transition.

In our revisited non-extensive, but thermodynamically consistent,
$q$-NJL model, we can perform calculations for both $q < 1$ and $q
> 1$ statistical effects. Some comments regarding their possible roles
are therefore necessary. Note first that the main approximation in
the NJL mean field theory description concerns the phase with
condensates, where the quarks are not confined but rather become
less massive approaching the phase transition region and
practically massless when crossing it (in the same way as in our
previous version of the $q$-NJL model, cf. Fig. 1 in
\cite{RWJPG}). In the usual NJL mean field model confinement is
not present; it can be introduced by adding a dynamical gluon
field, for example, in the form of a Polyakov loop \cite{PLoop,PLoop1,PLoop2}.
On the other hand, confinement of quarks in nucleons introduces
restrictions in the allowed phase space by changing their Fermi
motion and decreasing the chemical potential. It increases quark
arrangement in nuclear matter and, consequently, decreases the
entropy and effectively also the pressure between such composite
objects as nucleons. Such effects are visible in our $q$-NJL model
with $q < 1$. This means that for the phase with vacuum
condensates the $q < 1$ description could adequately describe the
correlations which are responsible for the changes in the
arrangement of quarks and therefore model, to some extent, the
effect of confinement. These correlations will survive even in the
unconfined phase (i.e. entropy there remains lower than in other
cases). Also, as seen in Table \ref{Table_1}, the critical
temperature for $q < 1$ is higher ($T^{(b)}_{cr} = 70.5$ MeV),
whereas the corresponding entropy becomes smaller. This is
understandable because at lower entropy we need larger temperature
to convey the same amount of energy.

For the case of $q > 1$  (case $(a)$) the average single particle
energy (chemical potential) is shifted towards higher values and
the entropy and pressure are also higher, increasing the Fermi
energy. The critical temperature (both for case $(a)$ and case
$(c)$) remains practically the same as for the extensive case with
$q=1$. The increase of pressure may simulate some additional
repulsion (at short distances) between massive quarks constituting
nucleons (see, for example, the quark-meson coupling models
\cite{QMC,QMC1}). This means that, effectively, $q > 1$ could be
regarded as emerging from the increasing nonstatistical
fluctuations found in the decay of dense fireballs produced in
heavy ion collisions and in other multiparticle production
experiments
\cite{multipart,multiparta,qMB,multipart1,multipart1a,multipartB,mpo,mpo1,mpo2,mpo3,mpo4,mpo5,CW,CWa}.

Summarizing, the nonextensive description accounts (in a
phenomenological way) for all situations in which one expects some
dynamical correlations in the quark-antiquark system considered in
our $q$-NJL model. It must be stressed that the nonextensive
approach is not a substitute for any part of the interaction
described already by the lagrangian of the NJL model, cf. Eq.
(\ref{lagr_eff}). It rather provides a different environment which
can have some dynamical effects, so far undisclosed but simply
parameterized by nonextensivity $q$. From this perspective case
$(b)$ with $q<1$, which has lower entropy, seems to be more
suitable to describe the nonextensive mechanism in the Equation of
State (EoS) of dense hadronic matter in a restricted phase space,
like, for example, in (proto)neutron stars. Respectively, the case
$(a)$, with $q > 1$ and with higher entropy, is more suitable for
situations in which one expects dynamical fluctuations, for
example to describe heavy ion collision where dense nuclear matter
is created  out of equilibrium and quickly decays into
hadrons\footnote{This fully agrees with the expected meaning of
the parameter $q$ discussed previously (in \cite{qcorrel,qcorrel1,qcorrel2} for $q <
1$ and in \cite{qfluct,qfluct1,qfluct2} for $q > 1$).}. Our work therefore
presents arguments that the critical properties of nuclear matter
in two different environments  can be different, although the
phase transition occurs at the same density or compression. The
critical temperature is higher for nuclear matter created in
(proto)neutron stars but the critical value of the chemical
potential will be bigger for nuclear matter created in heavy ion
collisions\footnote{The application of the formalism presented
here to calculations of the EoS for neutron matter with some
admixture of protons, including also hyperons for higher
densities, and to estimate the upper limit for the mass of neutron
stars \cite{stars,stars1} is underway. It is intended to continue recent
investigations presented in \cite{LPSTAR,AD_stars}.}

We conclude with two remarks. First: our approach should not be
confounded with the similar in spirit approach based on quantum
algebras (or on the so called $q$-deformed algebras) which was
used to formulate a $q$-deformed NJL model \cite{QA}. Their common
feature is the use of some suitable deformation of the mean field
NJL model (based on nonextensive statistical mechanics in our case
and on quantum algebras in \cite{QA}), which may account for
intrinsic correlations and fluctuations that go beyond the mean
field formulation and, in a certain limit, approach the more
realistic lattice calculations. Second: there exists another,
potentially very interesting, approach to nonextensivity, based on
the so called Kaniadakis entropy \cite{KanS,KanS1,KanS2}. In \cite{QCD_kappa}
it was used to study the formation of the quark-gluon plasma
formation (and compared with nonextensive approach) whereas in
\cite{Walecka_kappa} it was used to investigate the relativistic
nuclear EoS in the context of the Walecka quantum hadrodynamics
theory.

\vspace{0.3cm}

\noindent {\bf Acknowledgments}

\vspace{3mm}

The research of GW was supported in part by the National Science
Center (NCN) under contract Nr 2013 /08/M /ST2 /00598 (Polish
agency). We would like to thank warmly Dr Nicholas Keeley for
reading the manuscript.

\end{document}